\documentclass[prb,superscriptaddress,twocolumn]{revtex4-1}

\usepackage{graphicx,color}
\usepackage[caption=false]{subfig}
\begin{document}


\title{ Robust Intrinsic Ferromagnetism and Half Semiconductivity  in Stable Two-Dimensional Single-Layer Chromium Trihalides}

\author{Wei-Bing Zhang}
\email{Corresponding author. zhangwb@csust.edu.cn}
\affiliation{School of Physics and Electronic Sciences, Changsha University of Science and Technology, Changsha 410004, People's Republic of China}
\affiliation{Department of Applied Physics, Hong Kong Polytechnic University, Hung Hom, Hong Kong}
\author{Qian Qu}
\author{Peng Zhu}
\affiliation{School of Physics and Electronic Sciences, Changsha University of Science and Technology, Changsha 410004, People's Republic of China}
\author{Chi-Hang Lam}
\email{Corresponding author. C.H.Lam@polyu.edu.hk}
\affiliation{Department of Applied Physics, Hong Kong Polytechnic University, Hung Hom, Hong Kong}


\date{\today}

\begin{abstract}
Two-dimensional (2D) intrinsic ferromagnetic (FM) semiconductors are crucial to develop low-dimensional spintronic devices. Using density functional theory, we show that single-layer chromium trihalides (SLCTs) (CrX$_3$,X=F, Cl, Br and I)  constitute a series of  stable 2D intrinsic FM semiconductors. A free-standing SLCT can be easily exfoliated from the bulk crystal, due to a low cleavage energy and a high in-plane stiffness. Electronic structure calculations using the HSE06 functional indicate that both bulk and single-layer CrX$_3$ are half semiconductors with indirect gaps and their valence bands and conduction bands are fully spin-polarized in the same spin direction. The energy gaps and absorption edges of CrBr$_3$ and CrI$_3$ are found to be in the visible frequency range, which implies possible opt-electronic applications. Furthermore, SLCTs are found to possess a large magnetic moment of 3$\mu_B$ per formula unit and a sizable magnetic anisotropy energy. The magnetic exchange constants of SLCTs are then extracted using the Heisenberg spin Hamiltonian and the microscopic origins of the various exchange interactions are analyzed. A competition between a near 90$^\circ$ FM superexchange and a direct antiferromagnetic (AFM) exchange results in a FM nearest-neighbour exchange interaction. The next and third nearest-neighbour exchange interactions are found to be FM and AFM respectively and this can be understood by the angle-dependent extended Cr-X-X-Cr superexchange interaction.  Moreover, the Curie temperatures of SLCTs are also predicted using Monte Carlo simulations and the values can further increase by applying a biaxial tensile strain.  The unique combination of robust intrinsic ferromagnetism, half semiconductivity and large magnetic anisotropy energies renders the SLCTs as promising candidates for next-generation semiconductor spintronic applications.

\end{abstract}


\maketitle

\section{\label{intruduction}Introduction}
The  successful isolation of graphene from graphite has triggered extensive experimental and theoretical studies on two-dimensional (2D) materials.\cite{graphene_2004} Graphene has been found to possess many extraordinary properties including exceptionally high electron mobility ( $\sim$ 10$^5$ cm$^2$V$^{-1}$s$^{-1}$) and  Young's modulus ($\sim $ 1 TPa). \cite{roadmap} Besides graphene, various 2D materials including hexagonal boron nitride,\cite{hbn} silicene \cite{silicene} and transition metal dichalcogenides \cite{PhysRevLett.105.136805,nn} such as MoS$_2$ have also been identified as important candidates for  next-generation electronic and optoelectronic devices due to their unique  properties.\cite{pnas} Some available 2D materials may also be combined to produce novel van der Waals (vdW) heterostructures, which may lead to even richer physics and device applications.\cite{vdw}  However,  most of the available 2D materials  in the pristine form are intrinsically nonmagnetic, and this limits their applications in spintronics. Although magnetic moments can be introduced by dopants, defects, edges, and coupling to FM substrates \cite{spin}, a long-range magnetic order is rarely observed experimentally in 2D materials. It is thus desirable to explore novel 2D materials with a robust intrinsic ferromagnetic order.

Layered transition metal compounds constitute a large family of materials with a wide range of electronic, optical, and mechanical properties.\cite{nn,exf} The weak interlayer vdW bonding gives the possibility of their exfoliation into single-layer nanosheets.\cite{exf} Meanwhile, quantum confinement of electrons upon exfoliation is expected to lead to intriguing electronic,  optical and magnetic properties distinguished from their bulk. For example, bulk MoS$_2$ is an indirect semiconductor with a band gap of 1.2 eV, while single-layer MoS$_2$ has a direct band gap of 1.8 eV.\cite{PhysRevLett.105.136805,nn} Due to the direct band gap in the visible frequency range, MoS$_2$ is seen as an alternative to graphene and is considered as a promising candidate for  next-generation electronic and optoelectronic devices. With unfilled \emph{d/f} electrons in transition metal, the existence of magnetism is possible for transition metal compounds in both bulk and single-layer forms.  Recently, there is also increasing interest for exploring  novel single-layer transition metal compound  with long-range magnetic order.  Various 2D single-layer materials including K$_2$CuF$_4$ \cite{PRB_2d}, doped-MnPS$_3$ \cite{jacs_li} and CrSi(Ge)Te$_3$ \cite{jmc_li}  are predicted to be producible with an exfoliation process with the intrinsic intralayer ferromagnetic order of the bulk. Moreover, due to the strong spin-orbit coupling of transition metals, large magnetic anisotropy energies were also found in single-layer transition metal compounds \cite{LaMnAsO}, and this holds promise for low-dimensional magneto-electronics applications.

\begin{figure}
\centering
\includegraphics[width=0.35\textwidth]{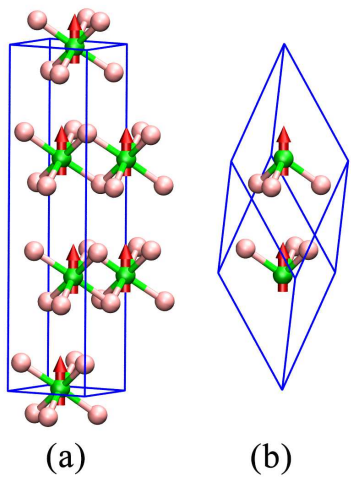}
\caption{\label{fig:bulk}(Color online) Crystal structure of ferromagnetic chromium trihalides CrX$_3$  in hexagonal (a) and rhombohedral (b) lattices. The red arrows represent the spin direction of Cr atoms. }
\end{figure}

In the present work, we focus on a series of layered semiconductors chromium trihalides, CrX$_3$(X=F, Cl, Br and I) \cite{crCl3,crBr3,crI3}. Although CrBr$_3$ in particular was known as the first ferromagnetic semiconductor found in 1960 \cite{jsps}, the knowledge of CrX$_3$ in general is still limited. It is known that  CrX$_3$(X=Cl, Br and I) crystalize in  the rhombohedral BiI$_3$ structure  (space group R$\overline{3}$ ) as shown in Fig.~\ref{fig:bulk} at low temperature and can be transformed into the monoclinic AlCl$_3$ structure (space group C2/m)  upon warming. In both phases, the Cr$^{3+}$ ions are arranged in a honeycomb network, and  coordinated by edge-sharing octahedra with six X$^-$ ions. The resulting  halide-Cr-halide triple layers are stacked  along the \emph{c}-axis with a van der Waals gap. All three compounds are found to exhibit a strong intralayer ferromagnetic coupling, but the interlayer couplings seem to be distinct.  CrCl$_3$ layers stacked antiferromagnetically with a N\'eel temperature ($T_N$) of 16.8 K \cite{Cable196129}, while CrBr$_3$  \cite{jsps}and CrI$_3$ \cite{crI3}layers are coupled ferromagnetically  with  Curie temperatures ($T_C$) of 37 K and 61 K. In addition,  the spins of the Cr atoms  are found to lie in the CrX$_3$  plane, whereas the easy axes  of the bromide and iodide are  found to be along the \emph{c}-axis. The combination of a weak interlayer coupling, an intrinsic magnetic order and semiconductivity of CrX$_3$ provide an excellent platform to explore the existence of novel 2D magnetic semiconductors, which are needed  by low-dimensional spintronics.   Very recently, McGuire \emph{et al.}  \cite{crI3} have re-examined the crystallographic and magnetic properties of CrI$_3$ experimentally, and also investigated both its bulk and single-layer properties using first-principles calculations. Their results indicate that  CrI$_3$ layers can be  easily exfoliated from bulk crystals and the ferromagnetic order is still retained  in the  single-layer form. However, detailed knowledge about the magnetic interactions and magnetic anisotropy of chromium trihalides, which is crucial to understand the magnetic mechanism and potential applications,  are still lacking.

Here, we present a comprehensive theoretical investigation on bulk chromium trihalides and their single-layer counterparts. The interlayer vdW interactions are included using the optB88-vdW functional, while the HSE06 functional is also used to obtain  accurate electronic structures and optical properties. The phonon dispersion calculations  and \emph{ab initio} molecular dynamics (MD) simulations are also performed to evaluate the stability of single layer.  Moreover, the magnetic exchange constants are extracted by comparing the total energy with the Heisenberg spin Hamiltonian. The Curie temperature is also predicted using Monte Carlo simulations. The microscopic origins of various exchange interactions involved are discussed in details.  The obtained geometries, electronic structures and magnetic properties are then compared with available experimental results. Our calculations show that single-layer chromium trihalides are stable two-dimensional intrinsic ferromagnetic half-semiconductors with large magnetic anisotropy energies. These make them promising materials for future spintronic applications.

\section{\label{comp}Computational Details}
The present calculations have been performed with spin-polarization using the Vienna ab initio simulation package (VASP) code \citep{vasp_CMS,vasp_PRB} within projector augmented-wave (PAW) method \cite{vasp_PAW,vasppawprb}.  General gradient approximations (GGA) in the Perdew-Burke-Ernzerhof(PBE) implementation \cite{PBE} are applied as the exchange correlation functional. To better take into account the interlayer van der Waals forces of  CrX$_3$, we also perform calculations adopting non-local van der Waals density functional in form of optB88-vdW \cite{optb88_prb,optb88_jpcm},  which  has been evidenced to give a much improved descriptions for various systems including graphene/metal interfaces.\cite{jcp}The range-separated hybrid functional HSE06 \cite{hse06} is used in calculating the electronic structure and optical properties accurately. In the HSE06 method,  the short-ranged exchange is constructed by mixing 25 \%  exact nonlocal Hartree-Fock exchange and  75 \%  semi-local PBE exchange. While both the electron correlation and long-ranged exchange are still treated at the PBE level.  Convergence tests have been performed carefully both for plane-wave cutoff energy and \emph{k}-point sampling. A plane-wave basis set expanded in energy with a cutoff of 600 eV is used in the calculation. \emph{K}-points sampling with a mesh of  8$\times$8$\times$8 \emph{k}-points generated by the scheme of Monkhorst-Pack \cite{VASP_MP} is used  for a primitive cell of bulk  CrX$_3$ as shown in Fig.~\ref{fig:bulk}-(b). And a 8$\times$8$\times$1 mesh is used for a single-layer  CrX$_3$ (Fig.~\ref{fig:mono}). All the lattice constants and atomic coordinates are optimized  until  the maximum force on all atoms is less than $5\times10^{-3}$  eV/\AA. A large vacuum space of at least 18 \AA~ thick is included in the supercell to avoid interaction between images. The elastic constants are obtained  by an energy-strain approach, in which  deformations applied to the unit cell and the corresponding  energy-strain relationship can be found in Ref.~\cite{PhysRevB.82.235414}.  The phonon calculations have been performed using the finite displacement approach, as implemented in the Phonopy code \cite{phonopy}, in which a 3$\times$3$\times$1 supercell and a displacement of 0.01 \AA~ from the equilibrium atomic positions are employed.

To extract the exchange interactions parameters between spins, we fit the total energies from DFT calculations for various spin configurations to  the Heisenberg spin Hamiltonian
\begin{equation}
\label{eq1}{
H=- J_1 \sum \limits_{\langle ij\rangle} \vec{S_i}\cdot\vec{S_j}- J_2\sum \limits_{\langle \langle ij\rangle \rangle} \vec{S_i}\cdot\vec{S_j}- J_3\sum \limits_{\langle \langle \langle ij\rangle \rangle \rangle} \vec{S_i}\cdot\vec{S_j}}
\end{equation}
defined on a honeycomb lattice. Here, $|\vec{S}|$=3/2.
Using these DFT-derived magnetic exchange parameters, the Curie temperature is then estimated using Metropolis Monte Carlo (MC) simulations, in which  a 100 $\times$100 $\times$ 1 2D honeycomb lattice with periodic boundary conditions is used. For each temperature studied, the MC simulation involves for 10$^5$ MC steps per site to attain thermal equilibrium. The temperature-dependent  mean magnetic moment and magnetic susceptibility are then obtained. The temperature at which the mean magnetic moment drastically drops to nearly zero and the magnetic susceptibility peaks sharply is identified as  the Curie temperature.

\section{\label{results}Results and Discussion}
\subsection{\label{sec:structure}Structure, Cleavage and Stability}
First, we discuss the structural properties of bulk chromium trihalides.   CrX$_3$ (X=Cl,Br and I) crystallize  in the R$\overline{3}$ phase at low temperature, and the corresponding hexagonal and rhombohedral lattices are shown in Fig.~\ref{fig:bulk}. It should be pointed out that CrF$_3$  is scarcely discussed in the literature. An experiment \cite{crF3} nevertheless indicates that it takes the R$\overline{3}$c phase. From experiments, the interlayer coupling is ferromagnetic for CrBr$_3$ and CrI$_3$, but antiferromagnetic for CrCl$_3$.  Both PBE and optB88-vdW calculations however indicate that the ferromagnetic state is more stable than the non-magnetic and anti-ferromagnetic phases in bulk for all four chromium trihalides. Similar results were also reported in early calculations.\cite{jpcm} Since the Cr atoms in the CrX$_3$ plane couple ferromagnetically in both ferromagnetic and antiferromagnetic  bulk CrX$_3$, the inconsistency of the interlayer magnetic state between experiment and theory for CrCl$_3$ should not effect the following calculation about single-layer chromium trihalides.  We thus only discuss the results of ferromagnetic CrX$_3$.

The lattice constants of the ferromagnetic CrX$_3$ calculated using the PBE and optB88-vdW functionals are given in Table.~\ref{tab:bulk}. Experimental results available from the literature are also listed for comparison. As shown in the table, both lattice parameters \emph{a} and \emph{c} increase with the element number of X. This can be attributed to the increasing atomic radius of X and the weakening reactivity between the X and the Cr atoms. It is clearly seen that the PBE functional significantly overestimates the interlayer distance $c$ by about 10\%. When the van der Waals forces are treated using the optB88-vdW functional, a good agreement between theory and experiment is achieved. This demonstrates that vdW forces play a very important role in interlayer binding and the optB88-vdW functional provides accurate  results. It will thus be used in calculating the cleavage energy of single-layer chromium trihalides.

\begin{table}
\centering
\caption{\label{tab:bulk}The calculated lattice parameters of CrX$_3$ using PBE and optB88-vdW functionals. Experimental results are also listed for comparison. For CrF$_3$, the lattice parameters  of R$\overline{3}$c phase are used. }
\begin{ruledtabular}
\begin{tabular}{ccccccccccc}
	&		&	PBE	&	optB88-vdW	&	experiment	\\
\hline
CrF$_3$	&	\emph{a}(\AA)	&	5.185 	&	5.163 	&	4.992\textsuperscript{\emph{a}} \\
	&	\emph{c}(\AA)	&	14.995 	&	13.351 	&	13.215 	\\
CrCl$_3$	&\emph{a}(\AA)	&	6.053 	&	5.987 	&	5.942\textsuperscript{\emph{b}}	\\
	&	\emph{c}(\AA)	&	19.422 	&	17.153 	&	17.333 	\\
CrBr$_3$	&	\emph{a}(\AA)	&	6.435 	&	6.350 	&	6.260\textsuperscript{\emph{c}}	\\
	&	\emph{c}(\AA)	&	20.836 	&	18.259 	&	18.200 	\\
CrI$_3$&	\emph{a}(\AA)	&	7.006 	&	6.905 	&	6.867\textsuperscript{\emph{d}}	\\
	&	\emph{c}(\AA)	&	22.350 	&	19.805 	&	19.807 	\\
\end{tabular}
\end{ruledtabular}
\textsuperscript{\emph{a}}Ref.~\cite{crF3}\\
\textsuperscript{\emph{b}}Ref.~\cite{crCl3}\\
\textsuperscript{\emph{c}}Ref.~\cite{crBr3}\\
\textsuperscript{\emph{d}}Ref.~\cite{crI3}\\
\end{table}

\begin{figure}
\centering
\includegraphics[width=0.45\textwidth]{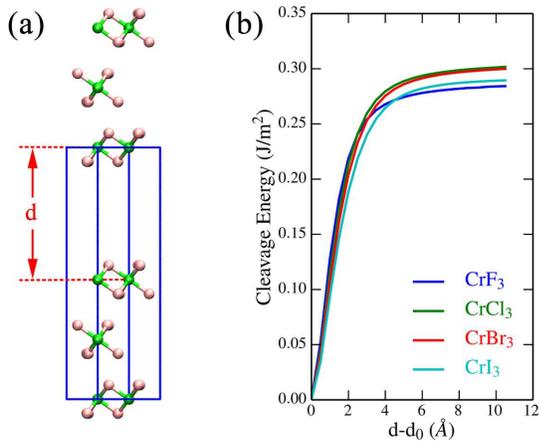}
\caption{\label{fig:cleavage}(Color online) Supercell model (a)  of  CrX$_3$ used to simulate the exfoliation procedure and the cleavage energy (b) calculated using the optB88-vdW functional as a function of the separation (d) between two fractured parts. d$_0$ represents the equilibrium interlayer distance of chromium trihalides. }
\end{figure}

We now focus on the possibility of the exfoliation of chromium trihalide layers from the bulk and the stability of free-standing single layers. To obtain a free-standing membrane, the cleavage energy needed to be overcome  in the exfoliation process should be small. In addition,  a high in-plane  stiffness is also needed to avoid curling upon exfoliation. \cite{PRB_2d}.  A large gap $d$ between two layers representing a fracture in the bulk as shown in Fig. \ref{fig:cleavage}-(a) is introduced to simulate the exfoliation procedure.  The cleavage energies as functions of \emph{d} calculated using the optB88-vdW functional are shown in Fig. \ref{fig:cleavage}-(b). The energy relative to the un-cleaved equilibrium state increases with the separation \emph{d}, and gradually converges to the cleavage energy. It can be seen that the cleavage energies for the four  CrX$_3$ materials are very similar and ranges only from 0.28 to 0.30  $J/m^2$. These values are comparable with other similar systems: CrSiTe$_3$ (0.35 $J/m^2$), CrGeTe$_3$ (0.38 $J/m^2$) \cite{jmc_li},  and MnPSe$_3$ (0.24 $J/m^2$) \cite{jacs_li}. They are also smaller than the experimentally estimated cleavage energy for graphite (0.36 $J/m^2$).  This indicates that the exfoliation of single layers from  bulk chromium trihalides is feasible experimentally. The theoretical cleavage strength given by the maximum derivative of the cleavage energy is 1.495, 1.384 ,1.303 and  1.177 GPa for CrF$_3$,CrCl$_3$, CrBr$_3$ and CrI$_3$, respectively.

\begin{figure}
\centering
\includegraphics[width=0.45\textwidth]{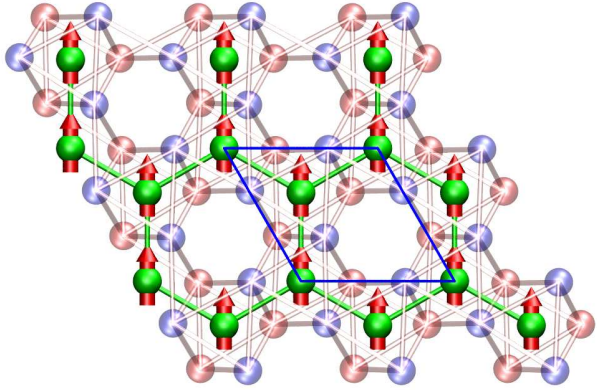}
\caption{\label{fig:mono}(Color online) The atomic structure and unit cell of ferromagnetic single-layer chromium trihalides upon exfoliation used in the electronic and phonon calculation. The transparent blue and red atoms represent the X atoms in top and bottom layer.  }
\end{figure}

In addition, the in-plane elastic constants  of single-layer chromium trihalides are also calculated using the PBE functional. The primitive cell of a 2D CrX$_3$ layer used in the calculations is shown in Fig.~\ref{fig:mono}. The calculated in-plane lattice constants as shown in Table.~\ref{tab:monolayer} are very similar to the bulk values.  Moreover, $c_{11}$, $c_{12}$, Young modulus Y$_{2D}$ and Poisson ratio $\nu$ decrease monotonically from F to I, which is also consistent with the binding strength between Cr$^{3+}$ and X$^-$. We further estimate the gravity induced out-of-plane deformation using the formula $h/L\approx \rho gL/Y_{2D}$ ,\cite{nll} where the length of the flake \emph{L} is set to 100 $\mu m$, and $\rho$ is the density of  2D chromium trihalides. The resulting values of $h/L$ for 2D CrX$_3$ (X=F,Cl,Br and I) are 3.405$\times$10$^{-4}$,3.629$\times$10$^{-4}$,	4.488$\times$10$^{-4}$, and 5.178$\times$10$^{-4}$. For graphene, $h/L$ is about $10^{-4}$ for flakes of this size. This suggests that free standing single-layer chromium trihalides are able to withstand their own weight and not curl easily. The low cleavage energy and strong in-plane stiffness promise that the 2D chromium trihalides can be produced by a simple exfoliation process.
\begin{table*}
\centering
\scriptsize
\caption{\label{tab:monolayer}The structural, elastic properties, cleavage strength($\sigma$)  magnetic anisotropy energy (MAE)  and easy axis (EA) of single-layer  chromium trihalides calculated using PBE functional.}
\begin{ruledtabular}
\begin{tabular}{ccccccccccc}
	&	\emph{a}(\AA)	&	$c_{11}$	(Nm$^{-1}$)&	$c_{12}$	(Nm$^{-1}$)&	Y$_{2D}$(Nm$^{-1}$)	&	$\nu$	&$\rho$(Kg m$^{-2}$)	&	$h/L$	&	$\sigma$ (GPa) &MAE ($\mu$eV/Cr)&EA	\\
\hline
CrF$_3$	&	5.189 	&	49.947 	&	23.862 	&	38.547 	&	0.478 	&	1.553$\times$10$^{-6}$	&	3.405$\times$10$^{-4}$	&	1.495&119.0&\emph{c}	\\
CrCl$_3$	&	6.051 	&	37.308 	&	11.069 	&	34.024 	&	0.297 	&	1.659$\times$10$^{-6}$	&	3.629$\times$10$^{-4}$	&	1.384 &31.5&\emph{c}	\\
CrBr$_3$	&	6.433 	&	31.741 	&	8.811 	&	29.295 	&	0.278 	&	2.703$\times$10$^{-6}$	&	4.488$\times$10$^{-4}$	&	1.303 &185.5&\emph{c}	\\
CrI$_3$	&	7.008 	&	25.474 	&	6.443 	&	23.845 	&	0.253 	&	3.378$\times$10$^{-6}$	&	5.178$\times$10$^{-4}$	&	1.177 &685.5&\emph{c}	\\
\end{tabular}
\end{ruledtabular}
\end{table*}

\begin{figure}
\centering
\includegraphics[width=0.5\textwidth]{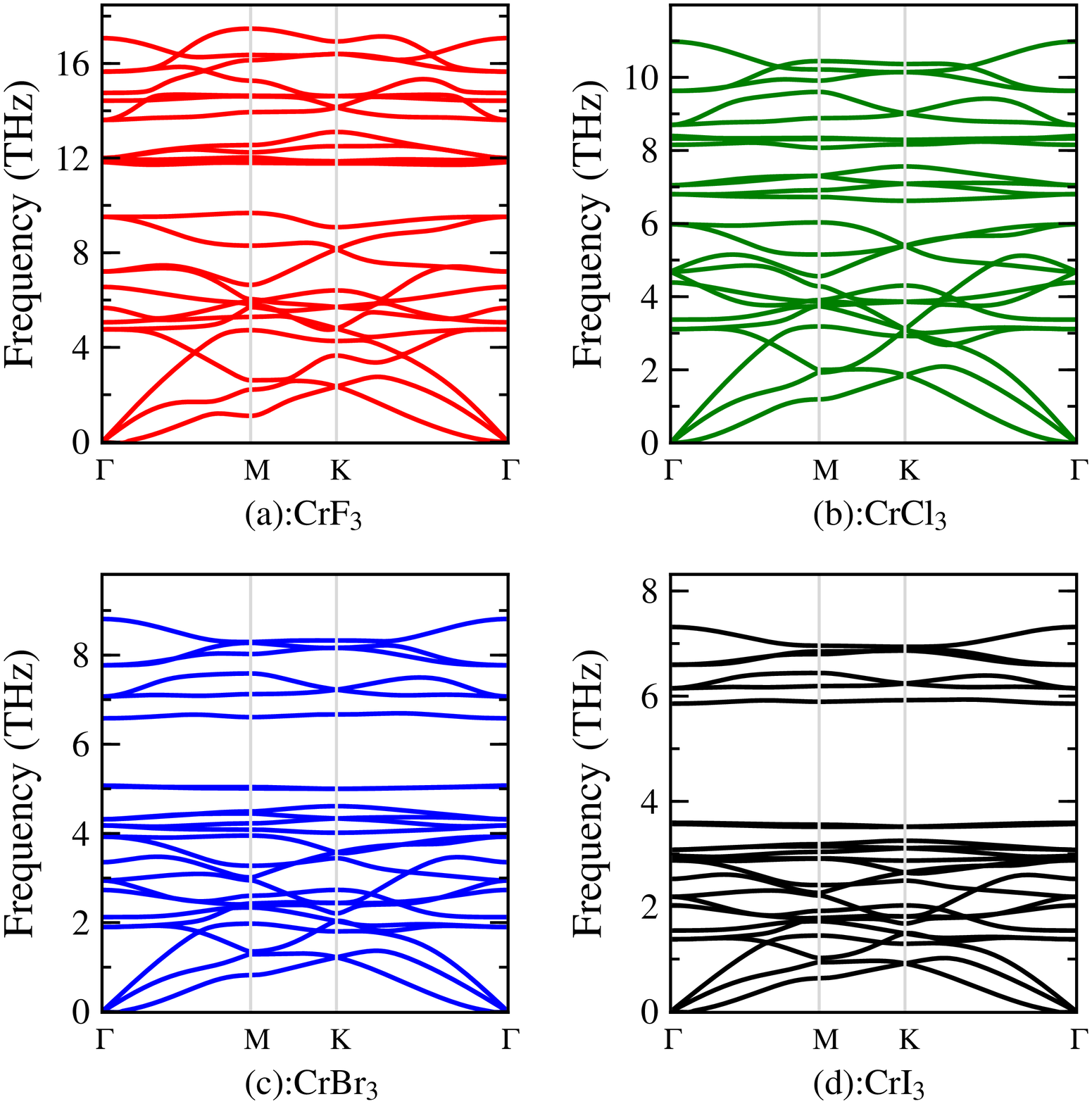}
\caption{\label{fig:phonon}(Color online)  Phonon dispersion of  single-layer chromium trihalides calculated using PBE functional.}
\end{figure}

\begin{figure}
\centering
\includegraphics[width=0.5\textwidth]{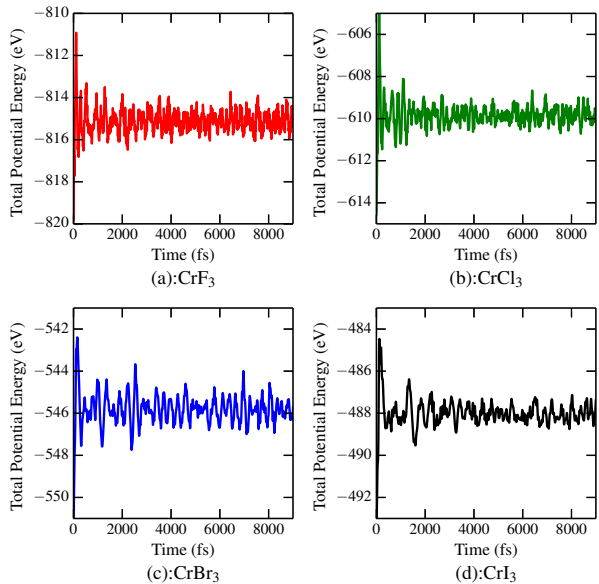}
\caption{\label{fig:md}(Color online)  Variations of the total potential energy of single-layer chromium trihalides with respect to simulation time during  \emph{ab initio} molecular dynamics  simulations. }
\end{figure}

To further confirm the stability of free-standing single-layer  chromium trihalides, we have calculated their phonon dispersion. As shown in Fig.~\ref{fig:phonon}, the imaginary frequency is found to be absent in the whole Brillouin Zone for all  2D chromium trihalides. This suggests that these single layer materials are dynamically stable and can exist as freestanding 2D crystals. In addition, the phonon frequency is found to be softening monotonically from F to I due to the increasing atomic weight of X. Furthermore,  we have performed  \emph{ab initio} molecular dynamics simulations at 300 K in canonical ensemble using Nos\'e heat bath scheme to examine the thermal stability of the single-layer chromium trihalides.  In these calculations, a large supercell of (4$\times$4) is used to minimize the constraint of  periodic boundary condition. Our results indicate that the atomic configurations of the single-layer chromium trihalides including the honeycomb network of the Cr$^{3+}$ ions and octahedra of CrX$_6$ remain nearly intact after heated for more than 9 ps with a time step of 3 fs. The variations of the total potential energy with respect to simulation time are plotted in Fig.~\ref{fig:md}. As shown in the figure, the total potential energies remain almost invariant during the simulation, which suggests that the single-layer chromium trihalides are thermally stable at room temperature. Clearly, the low cleavage energy, dynamics stability and thermal stability  of single-layer CrX$_3$ predicted here suggest that  the free-standing  ferromagnetic chromium trihalides  single layers can be realized experimentally even at room temperature.

\subsection{\label{sec:electronic}Electronic Structures and Optical properties}

\begin{figure*}
\centering
\subfloat[][]{%
\label{fig:band_ml_hse}%
\includegraphics[width=0.45\textwidth]{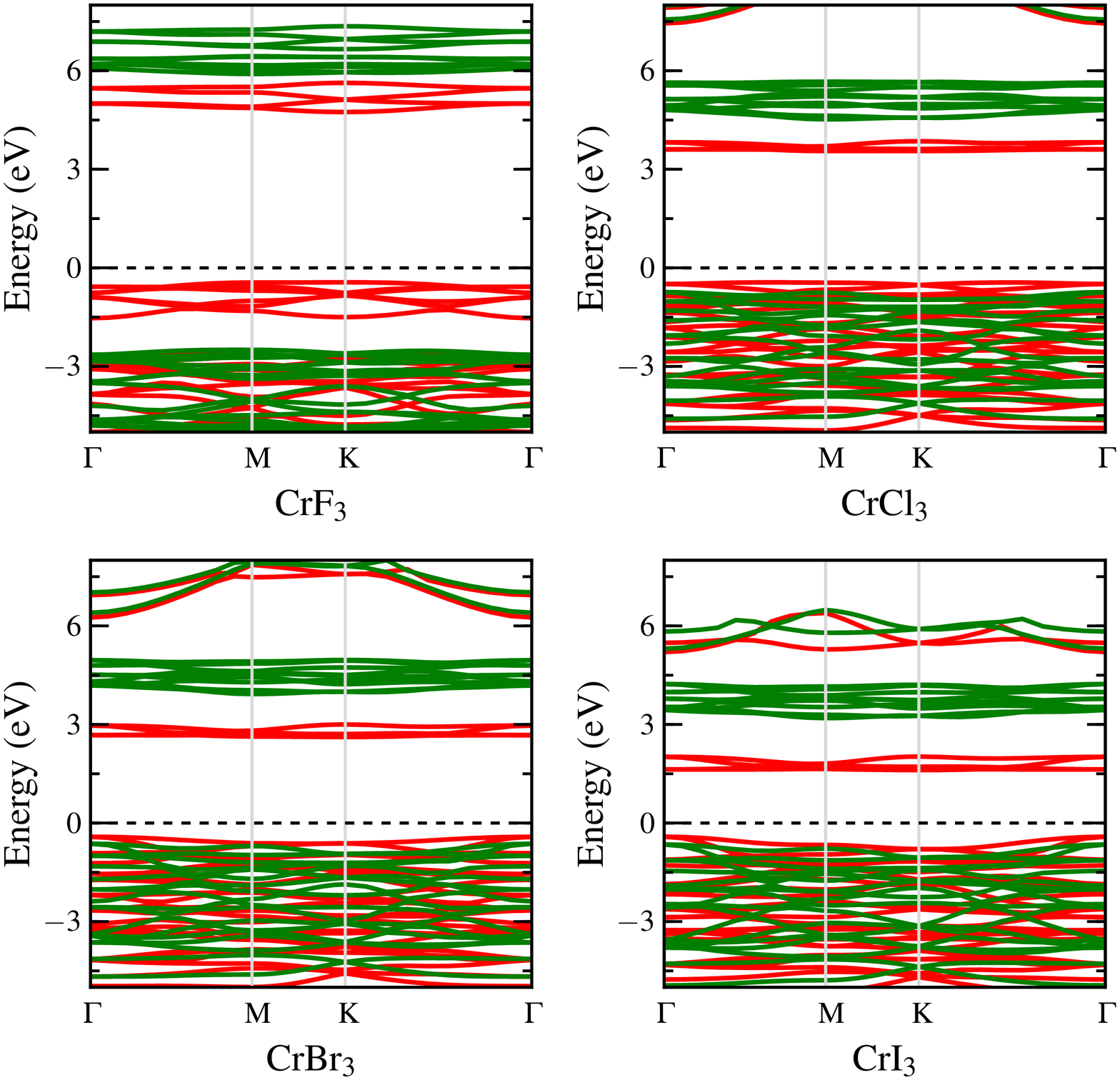}}
\subfloat[][]{%
\label{fig:band_ml_pbe}%
\includegraphics[width=0.45\textwidth]{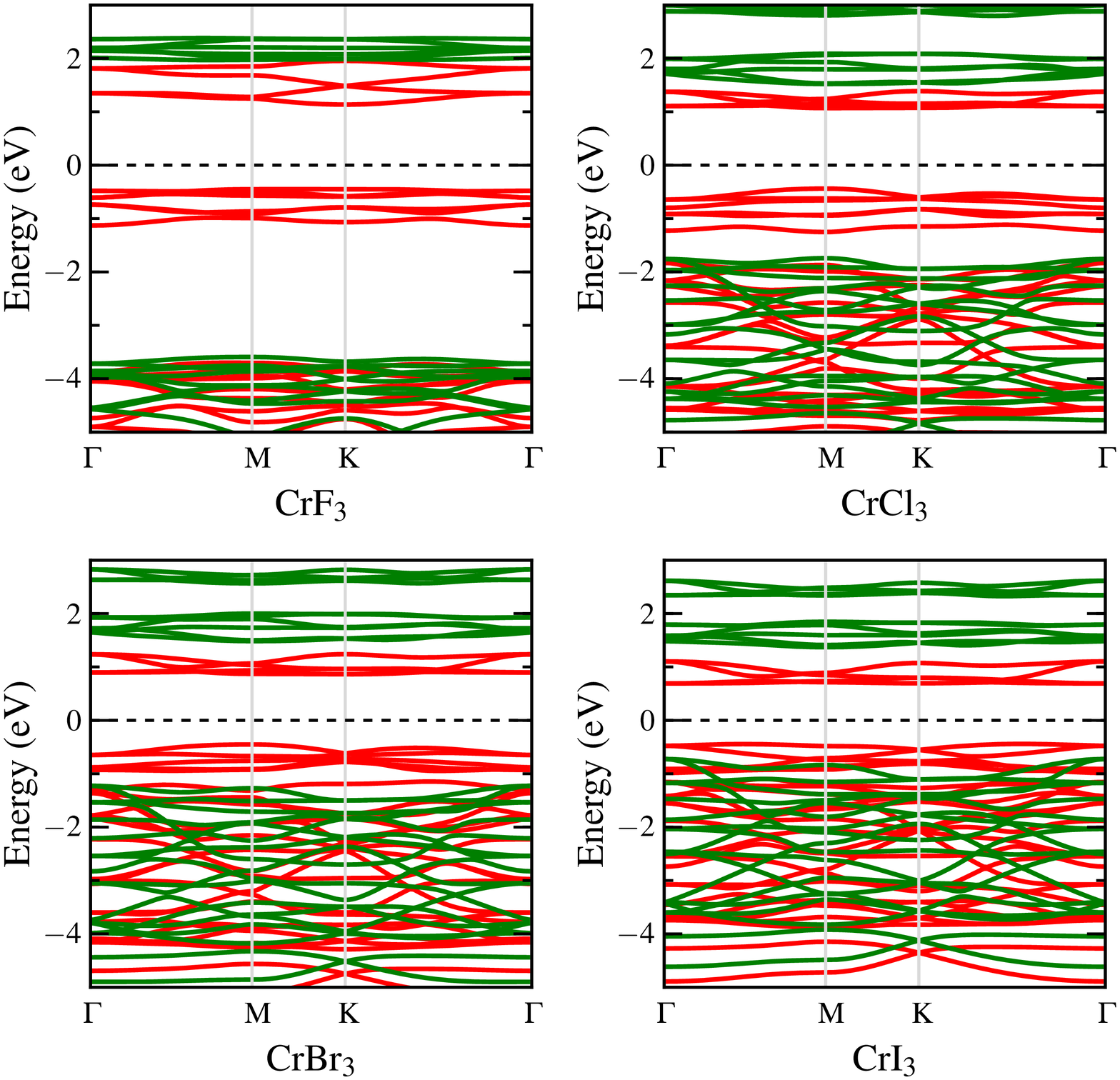}}
\caption{\label{fig:band_ML}(Color online) Band structures of  single-layer chromium trihalides calculated using HSE06 (a) and PBE (b) functionals. The red(blue) lines represent the band structure in the spin-up(spin-down) direction. }
\end{figure*}

\begin{figure*}
\centering
\subfloat[][]{%
\label{fig:dos_ml_hse}%
\includegraphics[width=0.45\textwidth]{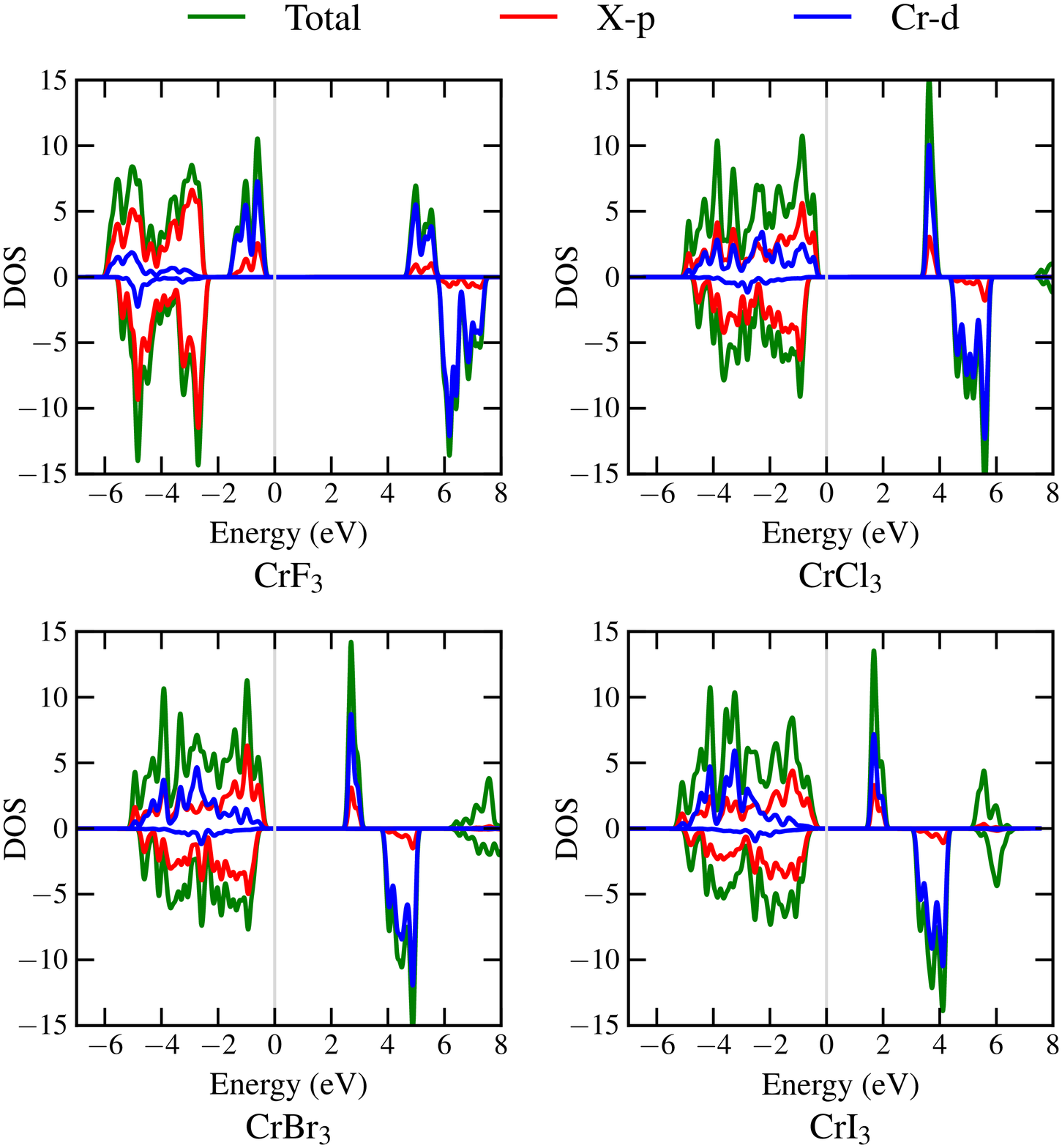}}
\subfloat[][]{%
\label{fig:dos_ml_pbe}%
\includegraphics[width=0.45\textwidth]{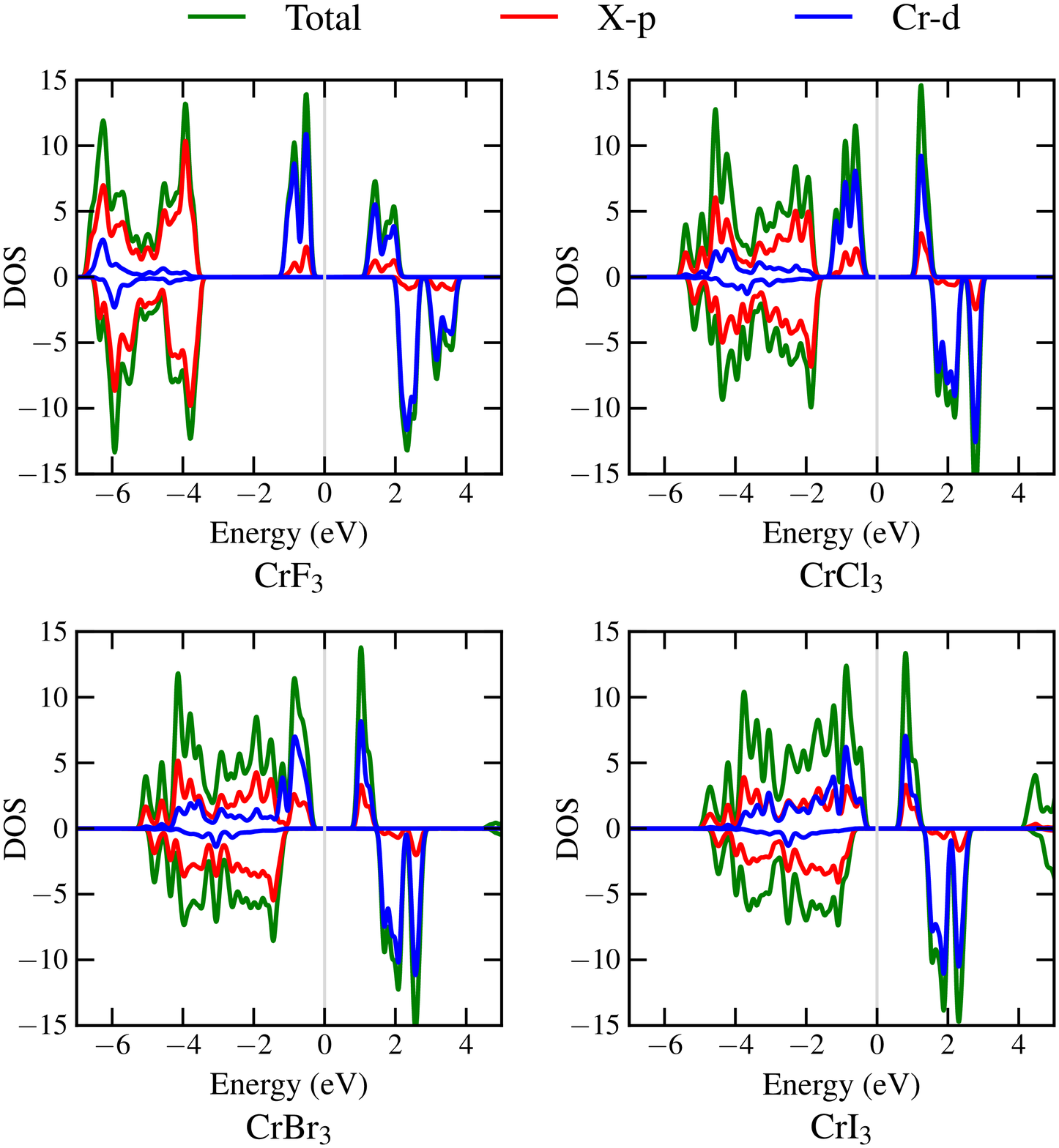}}
\caption{\label{fig:dos_ML}(Color online)  Total and partial density of states of single-layer chromium trihalides calculated using HSE06 (a) and PBE(b) functionals.}
\end{figure*}

\begin{table*}
\centering
\caption{\label{tab:ele_ml}The characteristic parameter in unit of eV of energy band  of single layer and bulk chromium trihalides calculated using PBE and HSE06 functionals.}
\footnotesize
\begin{ruledtabular}
\begin{tabular}{cccccccccccccccccccc}
	\multicolumn{9}{c}{PBE}		\\
	&		\multicolumn{4}{c}{bulk}	&\multicolumn{4}{c}{single layer}	&		\\
	&	$\Delta_{cbm}$	&	$\Delta_{vbm}$	&	Gap$_{up}$&Gap$_{down}$	&	$\Delta_{cbm}$	&	$\Delta_{vbm}$&	Gap$_{up}$&Gap$_{down}$		\\
CrF$_3$	&	0.816	&	-3.102	&	1.609 &	5.527	&	0.818	&	-3.146	&	1.614	&5.578	\\
CrCl$_3$	&	0.474	&	-1.324	&	1.454	&	3.252	&	0.451	&	-1.309	&	1.521&	3.281	\\
CrBr$_3$	&	0.62	&	-0.721	&	1.182 &	2.523	&	0.624	&	-0.776	&	1.331 &	2.731	\\
CrI$_3$	&	0.754	&	-0.231	&	0.913 &	1.898	&	0.686	&	-0.280	&	1.143&	2.109	\\
\\
\multicolumn{9}{c}{HSE06}		\\
	&		\multicolumn{4}{c}{bulk}	&\multicolumn{4}{c}{single layer}	&		\\
	&	$\Delta_{cbm}$	&	$\Delta_{vbm}$	&	Gap$_{up}$&Gap$_{down}$	&	$\Delta_{cbm}$	&	$\Delta_{vbm}$&	Gap$_{up}$&Gap$_{down}$		\\
CrF$_3$	&	1.075	&	-1.862	&	4.728 &	7.665	&	1.14	&	-2.071	&	4.680 &	7.891	\\
CrCl$_3$&	0.987	&	-0.267	&	3.305 &	4.559	&	0.992	&	-0.343	&	3.441	&	4.776\\
CrBr$_3$	&	1.316	&	-0.167	&	2.402&	3.885	&	1.322	&	-0.220	&	2.538&	4.080	\\
CrI$_3$&	1.639	&	-0.204	&	1.286 &	3.129	&	1.609	&	-0.243	&	1.525	&	3.377\\
\end{tabular}
\end{ruledtabular}
\end{table*}

We further investigate the electronic structures of single-layer chromium trihalides. Since PBE functional usually underestimates the energy band gap, hybrid functional in form of HSE06 functional is used to get accurate electronic structures.   As shown in Fig.~\ref{fig:band_ml_hse}, all  monolayer chromium trihalides  are indirect semiconductors. The band gap decreases with element number of X and the corresponding values are 4.68, 3.44, 2.54, and 1.53 eV for CrX$_3$(X=F,Cl,Br and I) respectively. Moreover, we also find from Fig.~\ref{fig:dos_ml_hse} that the valence band and conduction band edges around the Fermi level are fully spin-polarized and exclusively contributed by same spin component, which shows a typical half-semiconductor character. The difference of the band edge energy between the two spin components (denoted by $\Delta_{cbm}=E_{cbm}^{down}-E_{cbm}^{up}$ and  $\Delta_{vbm}=E_{vbm}^{down}-E_{vbm}^{up}$ for the conduction band minimum  and the valance band maximum respectively) are also reported in Tab.~\ref{tab:ele_ml}. In particular,  $\Delta_{cbm}$ reaches 1.609 eV in the case of CrI$_3$, which implies possible applications in spin-polarized carrier injection and detection. More interestingly,  occupied Cr-3d orbitals are only found in  the spin-up direction  and the  spin-down Cr-3d states are fully unoccupied. These results can be understood using a crystal field theory. Since the Cr$^{3+}$ ions are located at a  octahedra environment  coordinated by  six X$^-$ ions,  the Cr-3d orbitals are spitted into three low-lying t$_{2g}$ and  two high-lying e$_g$ orbitals. According to Hund's rule and the Pauli exclusion principle, Cr$^{3+}$ will take a high-spin t$_{2g}^{3}$ e$_{g}^{0}$ electronic configuration, which implies that the occupied 3d orbitals must all be in one of the spin directions. Moreover, we can also find interesting element- and spin-dependent gap character in CrX$_3$ . As shown in Fig.~\ref{fig:dos_ml_hse},  HSE06 calculations indicate that the conduction band is dominated by Cr-3d states weakly hybridized with the X-3p states for all four compounds in both spin directions. And the states of  the valence band in spin-down direction are of almost pure  X-3p  character, which is  also element-independent.  For the upper part of the valence band, the states are contributed by the X-3p states  with a  mixture of  Cr-3d states in CrX$_3$ (X=Cl,Br and I) and the contribution of X-3p increases with the element number of X.  However,  the Cr-3d states of  CrF$_3$  still  dominate the valence band near fermi level.   Hence  the band gaps  in spin-down direction predicted by HSE06 are of charge-transfer-type but shows a element-dependent behaviour in spin-up direction.  We can find that the gap is Mott-Hubbard type for  CrF$_3$ and will gradually change from a mixture of a Charge-Transfer and a Mott-Hubbard type for CrCl$_3$ and CrBr$_3$ to Charge-Transfer type in case of CrI$_3$. Results based on the PBE functional (Fig.~\ref{fig:band_ml_pbe} and Fig.~\ref{fig:dos_ml_pbe}) also implies indirect semiconductors  but with a much smaller band gap.  However, it should be noticed that HSE06 and PBE functionals give different hybridization behaviours. Compared with HSE06 results, the crystal field splitting and exchange splitting of the Cr-3d states are strongly deceased in PBE functional (Fig.~\ref{fig:dos_ml_pbe}). And the Cr-3d states (i.e. t$_{2g}$ states)  are shifted to higher energies closed to Fermi level  compared with the X-3p states and predominated the valence band maximum, resulting in a strongly increased  hybridization between X-3p and Cr-3d states.  PBE functional thus tends to give  a Mott-Hubbard type band gap  in spin-up direction whereas  the gap character  in spin-down direction is still same to HSE06. Further experiments should be conducted to clarify this interesting spin-dependent electronic structure of CrX$_3$.

\begin{figure*}
\centering
\subfloat[][]{%
\label{fig:optical_ml_hse}%
\includegraphics[width=0.45\textwidth]{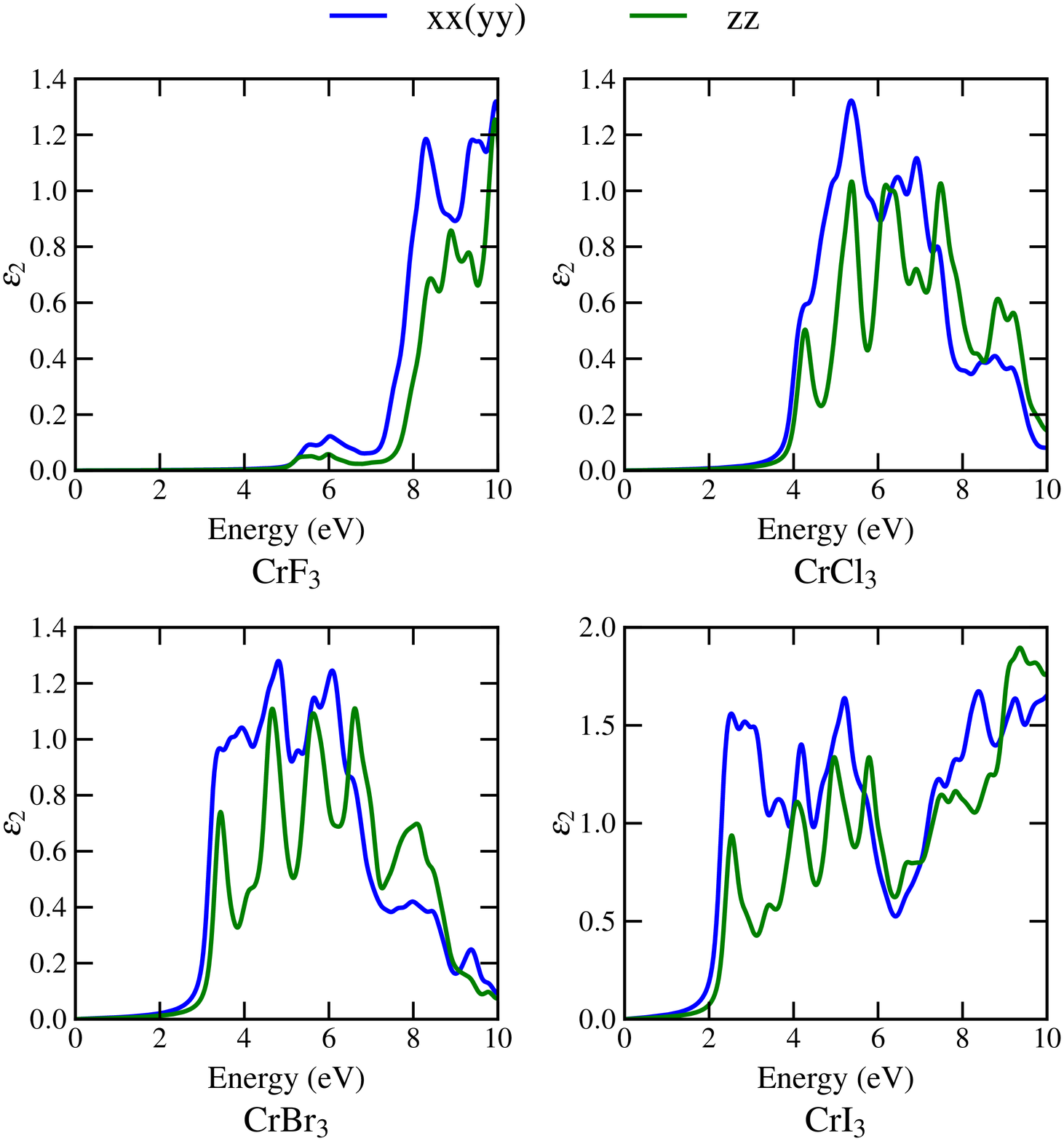}}
\subfloat[][]{%
\label{fig:optical_ml_pbe}%
\includegraphics[width=0.45\textwidth]{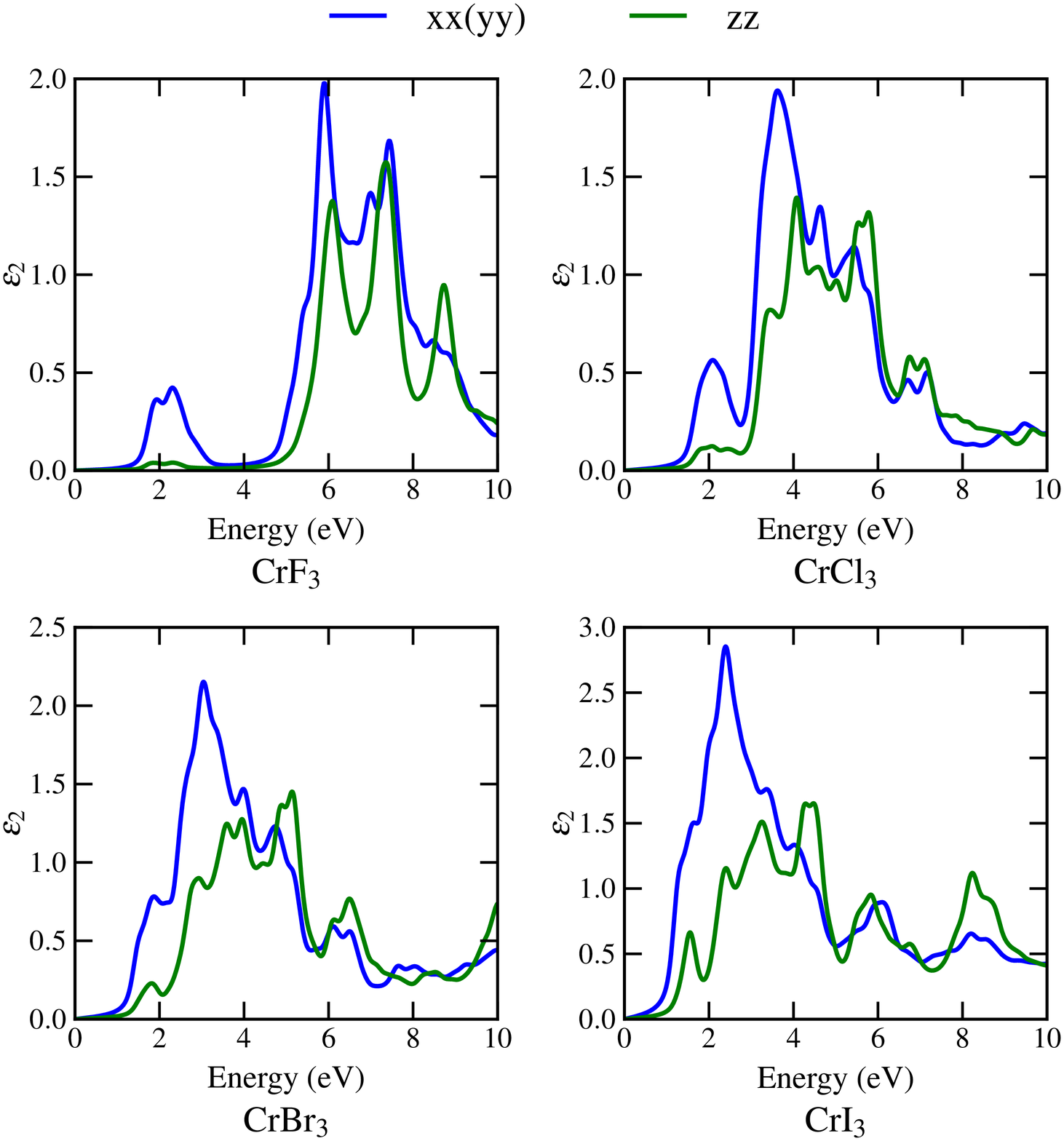}}
\caption{\label{fig:optical_ML}(Color online)  Imaginary part of the dielectric function $\epsilon_2$ of single-layer chromium trihalides calculated using  HSE06 (a) and PBE (b) functionals. }
\end{figure*}

We have also calculated the optical properties of CrX$_3$.  As shown in Fig.~\ref{fig:optical_ml_hse}, due to hexagonal symmetry of the CrX$_3$ lattice, the dielectric functions  along the \emph{x} and \emph{y} axis are identical  but different from that along the \emph{z} axis, implying anisotropic optical  properties. We can also find that  the main peak of  $\epsilon_2$  in CrX$_3$ can be traced back to  the p-d excitation from X-3p states of valence band to Cr-3d states of lowest conduction band. For CrF$_3$, the satellite states around 6 eV are contributed by  d-d excitation from occupied 3d states around -1 eV and unoccupied states around 5 eV as shown in Fig.~\ref{fig:dos_ml_hse}.  Our PBE calculations (Fig.~\ref{fig:optical_ml_pbe}) give similar results  but  the absorption edges at lower energies than those found using HSE06. In addition, the  stronger satellite states induced  by d-d excitation are found to exist in CrX$_3$, which can be understood by the fact that  the valence band maximum is predominated by Cr-3d states in PBE calculation. It should be noticed that the main absorption peaks for CrBr$_3$ and CrI$_3$  fall in the visible range, although they are indirect semiconductors.  This suggests that CrBr$_3$ and CrI$_3$ are possible candidate for optoelectronic application.

\begin{figure*}
\centering
\subfloat[][]{%
\label{fig:band_bulk_hse}%
\includegraphics[width=0.45\textwidth]{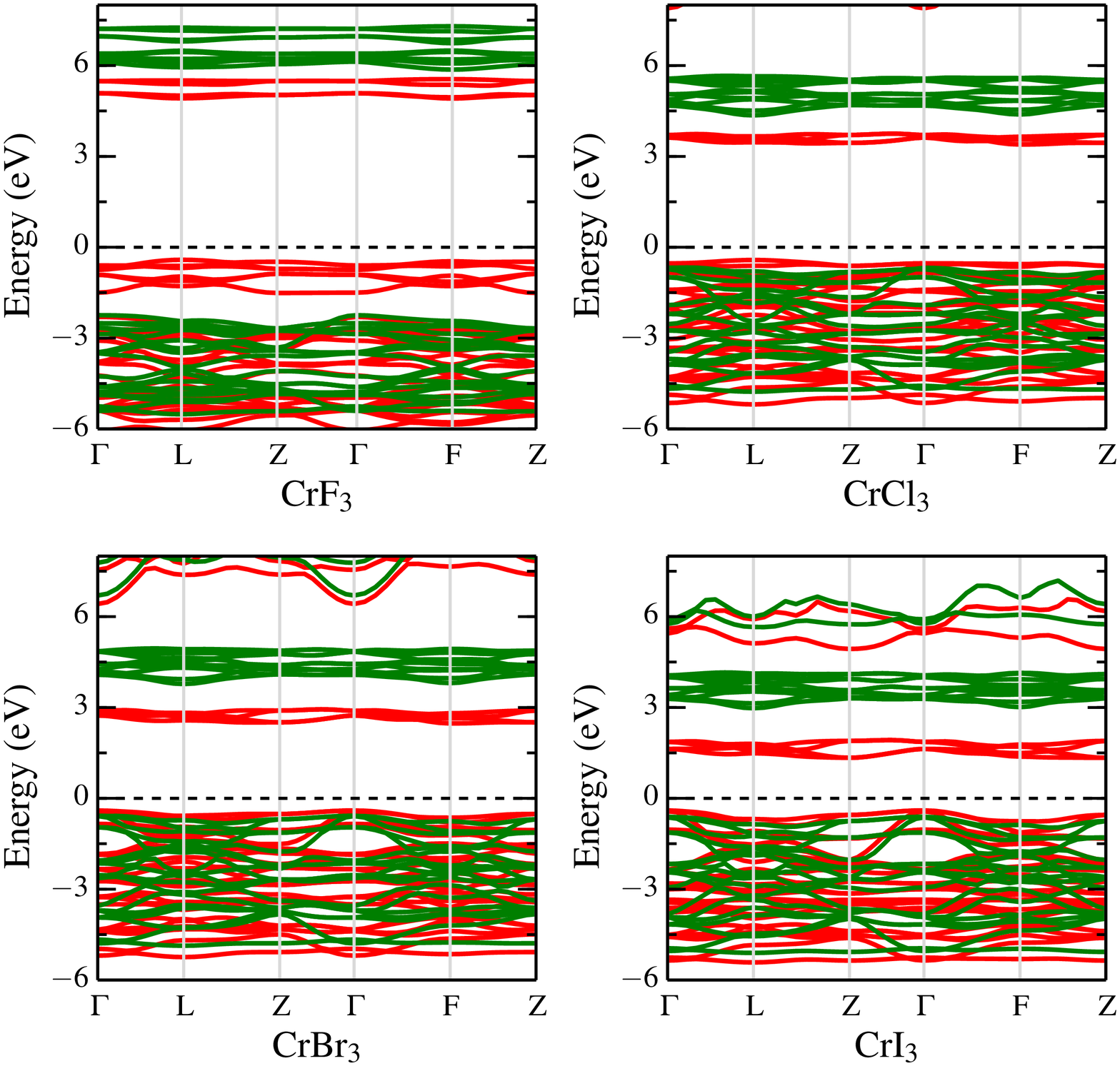}}
\subfloat[][]{%
\label{fig:band_bulk_pbe}%
\includegraphics[width=0.45\textwidth]{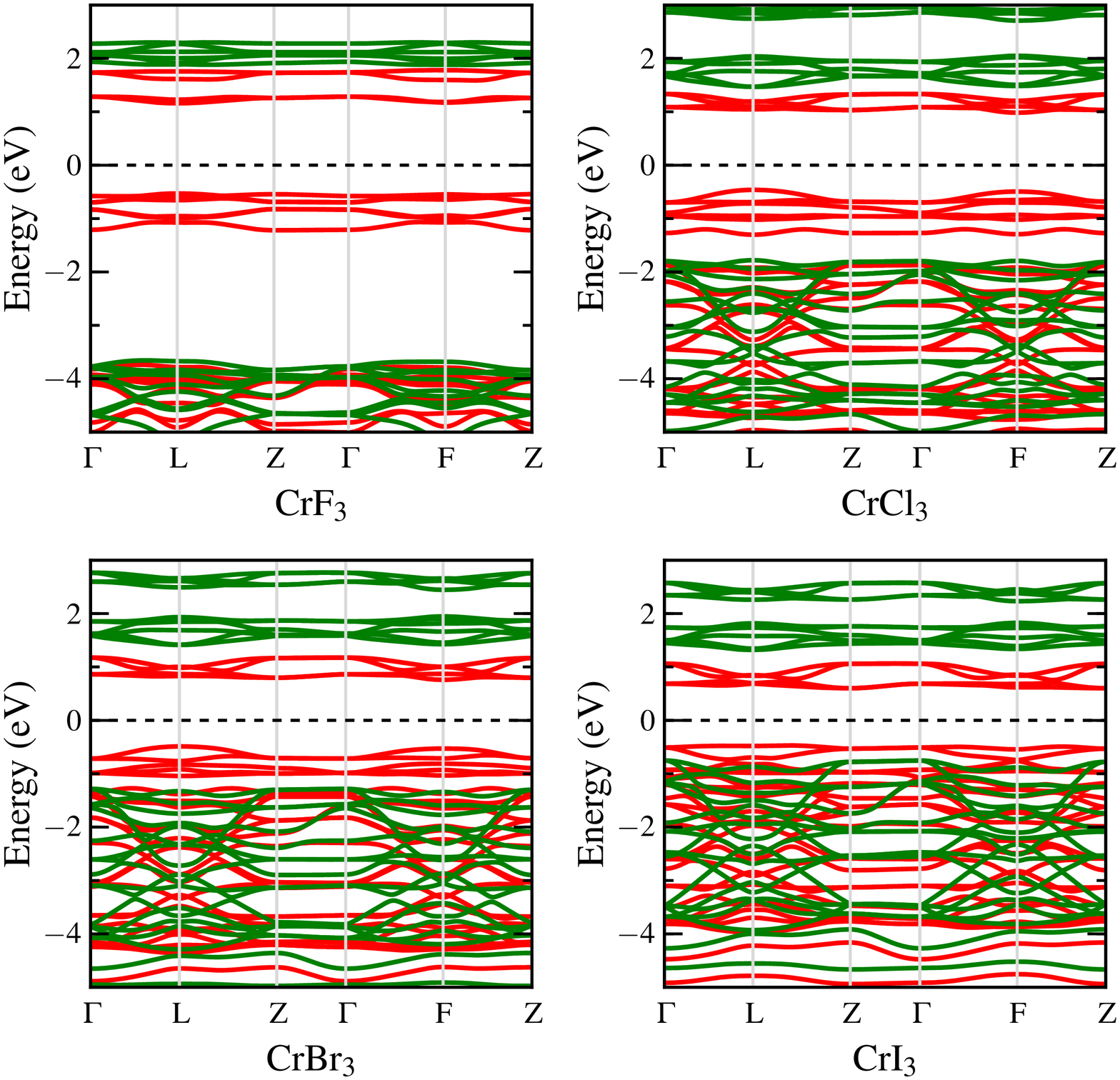}}
\caption{\label{fig:band_bulk}(Color online) Band structure of bulk chromium trihalides calculated using HSE06 (a) and PBE (b) functionals. The red(blue) lines represent the band structure in the spin-up(spin-down) direction.}
\end{figure*}

\begin{figure*}
\centering
\subfloat[][]{%
\label{fig:dos_bulk_hse}%
\includegraphics[width=0.45\textwidth]{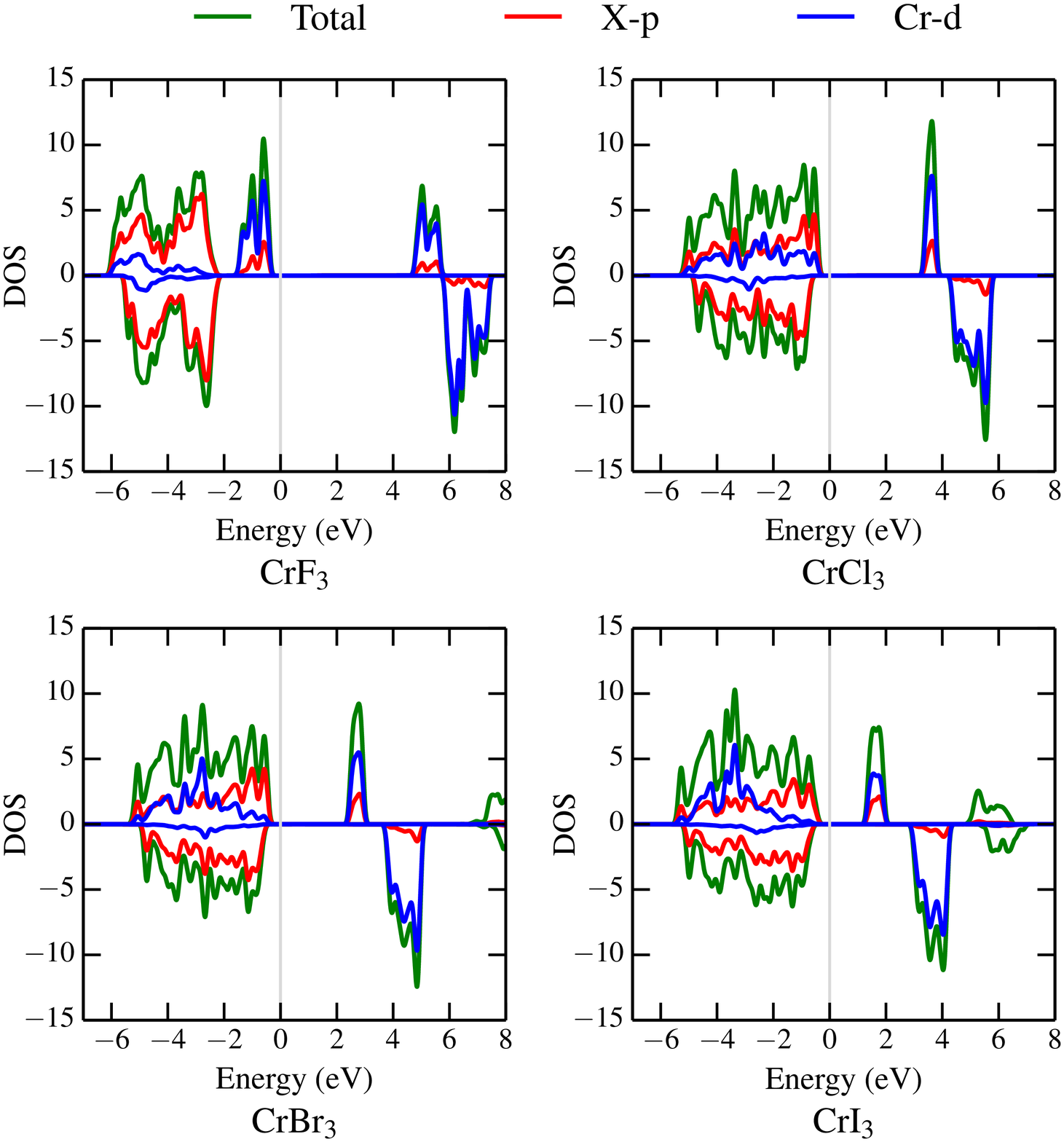}}
\subfloat[][]{%
\label{fig:dos_bulk_pbe}%
\includegraphics[width=0.45\textwidth]{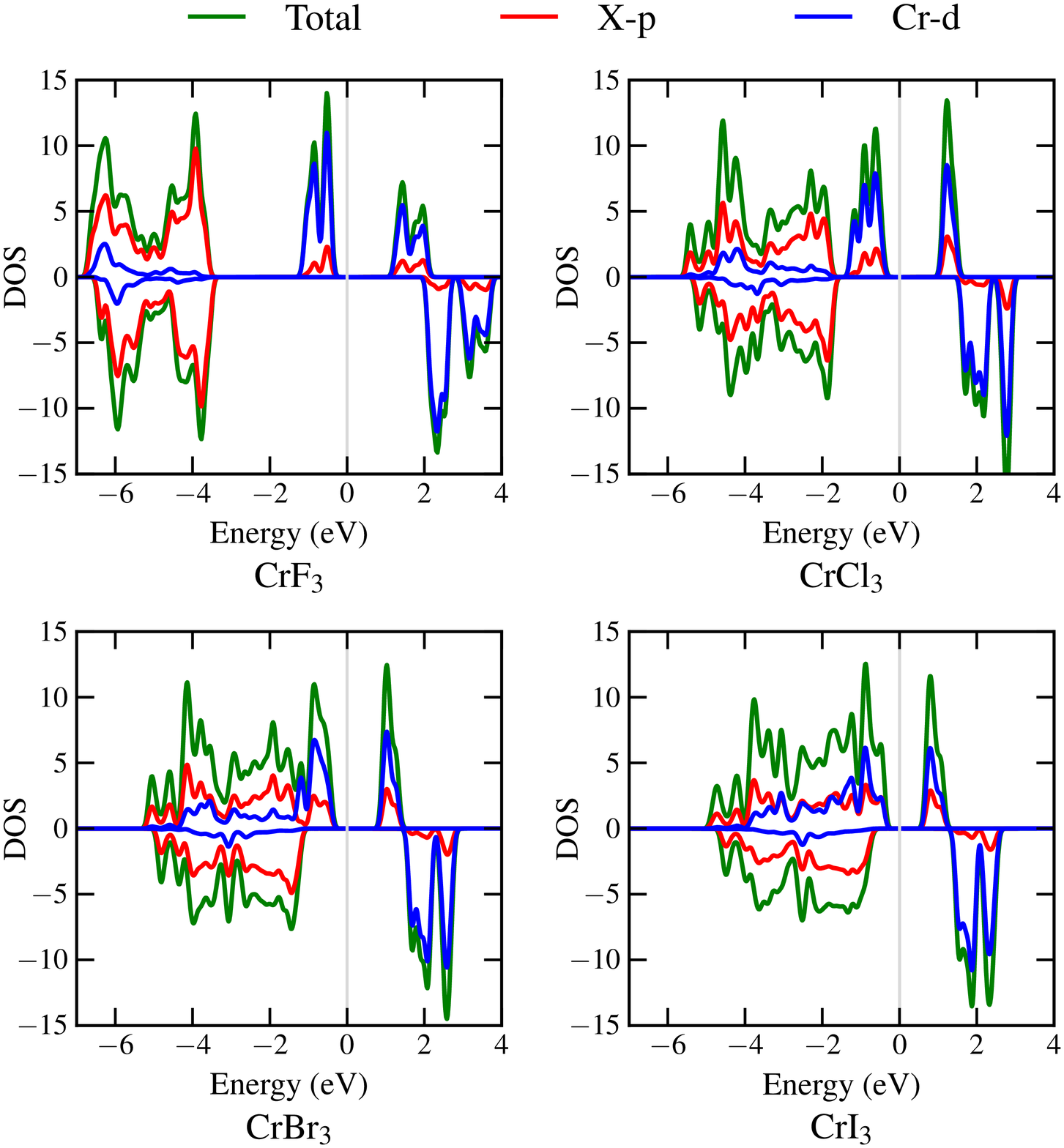}}
\caption{\label{fig:dos_bulk}(Color online) Total and partial density of states of bulk chromium trihalides calculated using HSE06 (a) and PBE (b) functionals.}
\end{figure*}

\begin{figure*}
\centering
\subfloat[][]{%
\label{fig:opt_bulk_hse}%
\includegraphics[width=0.45\textwidth]{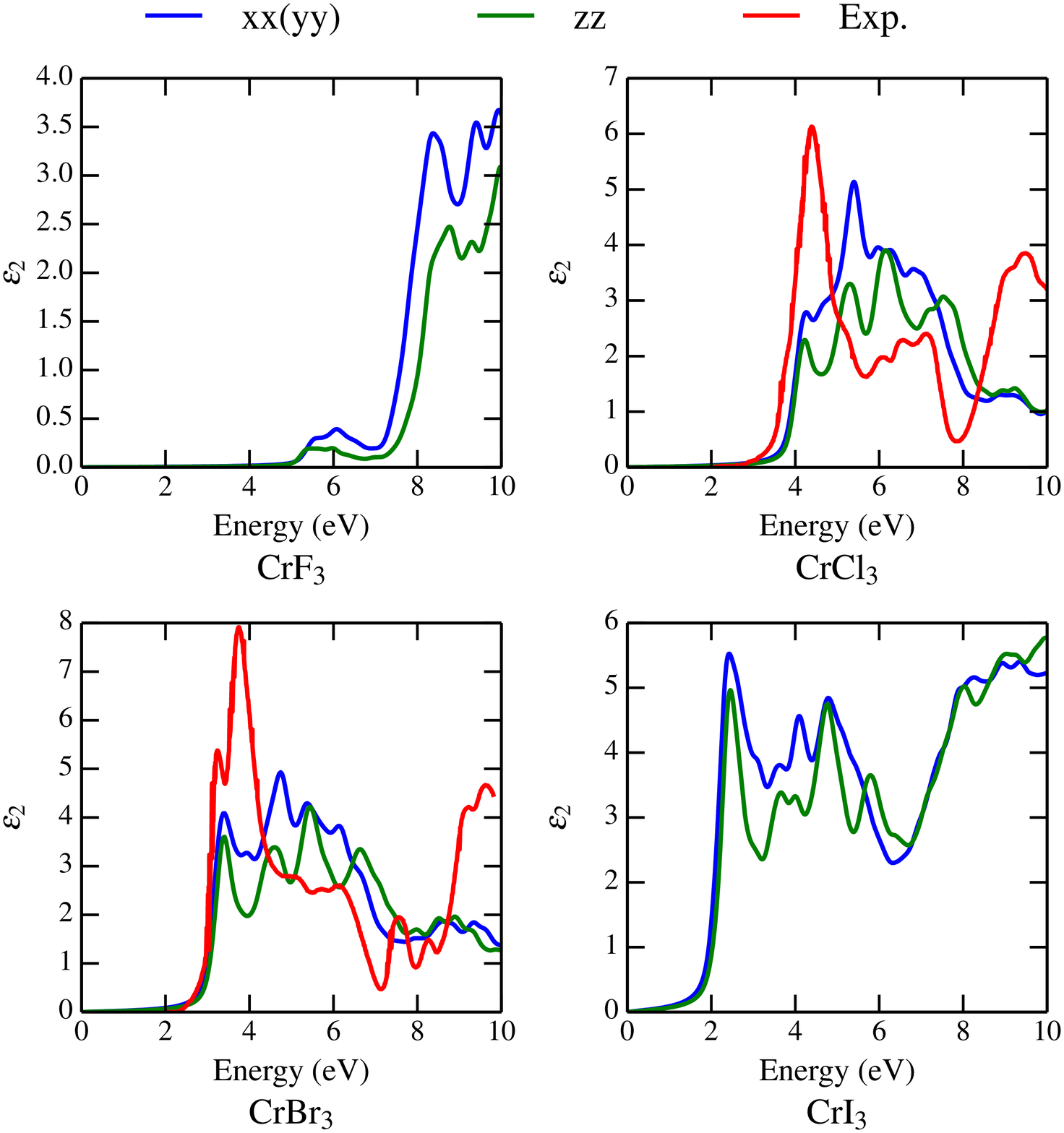}}
\subfloat[][]{%
\label{fig:opt_bulk_pbe}%
\includegraphics[width=0.45\textwidth]{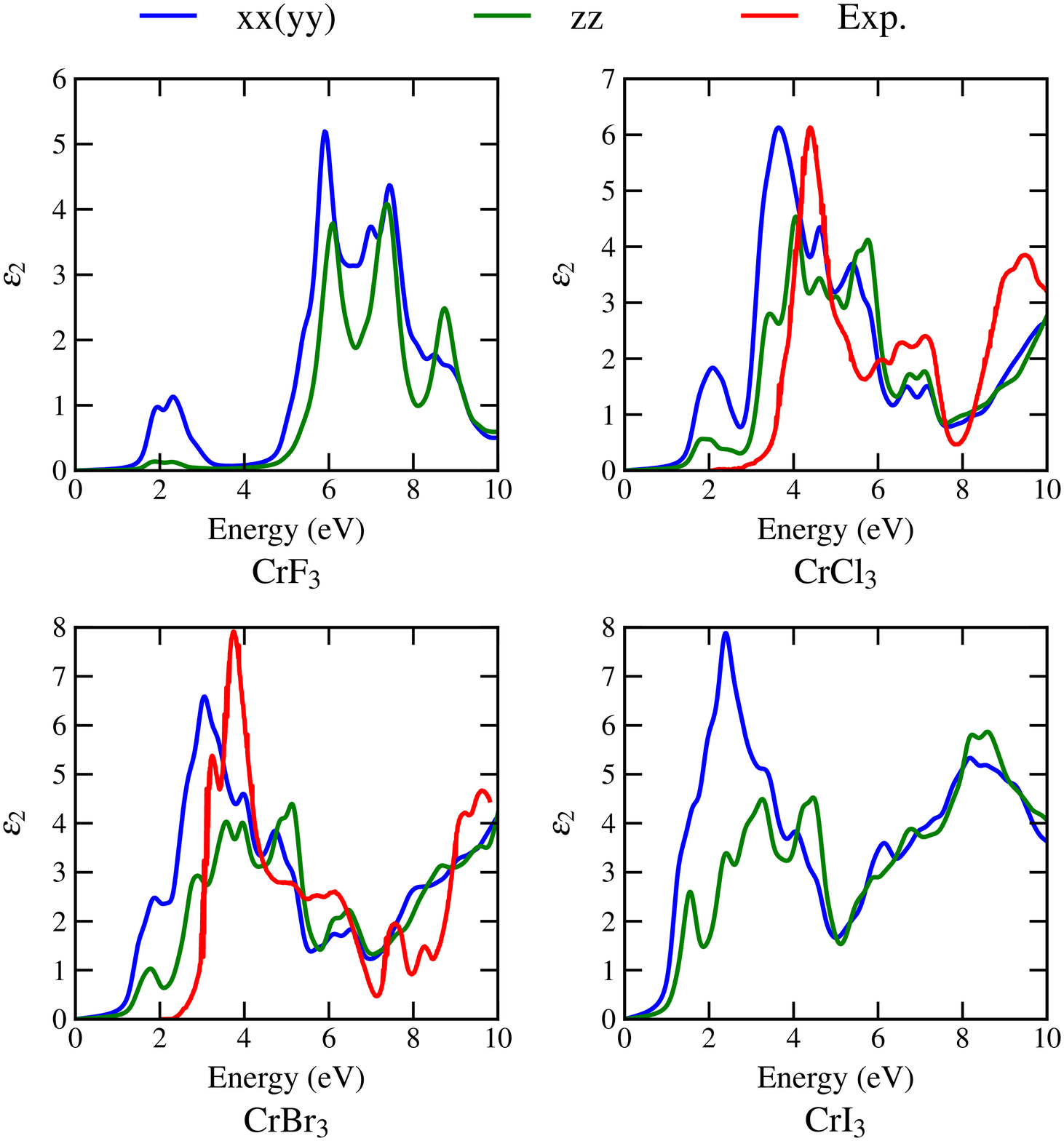}}
\caption{\label{fig:opt_bulk}(Color online) Imaginary part of the dielectric function of bulk chromium trihalides calculated using HSE06 (a) and PBE (b) functionals. Available experimental data from Ref.~\cite{opt_exp} are also shown for comparison. }
\end{figure*}

To reveal the change of electronic and optical properties of CrX$_3$  due to quantum confinement,  we have also calculated the electronic structure and optical properties of bulk CrX$_3$ using the HSE06 and PBE functionals. Our results indicate that the band structure (Fig.~\ref{fig:band_bulk}) and density of states (Fig.~\ref{fig:dos_bulk}) of bulk CrX$_3$  are very similar to their single layer counterparts. This is except that the band gap  is slightly smaller, which is also found in most 2D transition metal compounds.  Experimentally, the band gaps of  bulk CrBr$_3$ and CrI$_3$ are 2.1 eV \cite{gap_CrBr3} and 1.2 eV \cite{gap_CrI3} respectively, which are in good agreement with 2.40 and 1.28 eV given by our HSE06 calculations.  We also compare our calculated imaginary part of the dielectric function with available experimental results for  CrBr$_3$  and CrCl$_3$ \cite{opt_exp}. As shown in Fig.~\ref{fig:opt_bulk_pbe}, PBE underestimates the band gap significantly, and this leads to an  absorption edge at a lower energy than the experimental value. In addition, PBE also gives a satellite peak which seems to be invisible in experiment. The HSE06 calculations seem to give more accurately predictions of the optical properties and a good agreement between theory and  experiment is achieved in cases of  CrBr$_3$  and CrCl$_3$. This also verifies that our HSE06 calculations are reliable and are hence expected to give accurate band gaps and optical properties for single-layer chromium trihalides.

\subsection{\label{sec:magnetic}Magnetic Properties }

\begin{figure}
\centering
\includegraphics[width=0.45\textwidth]{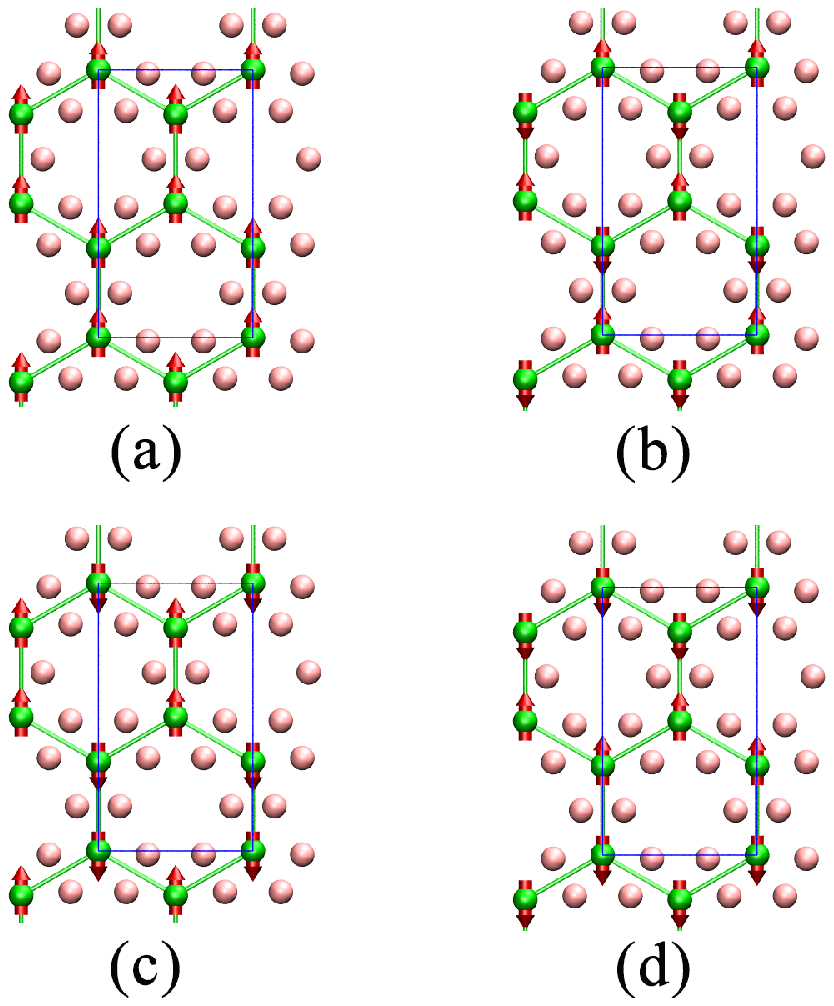}
\caption{\label{fig:magnetic}(Color online) The possible magnetic configuration of  single layer  chromium trihalides CrX$_3$: FM  (a), AF-N$\acute{e}$el (b), AF-zigzag (c), and AF-stripy ordered (d).  The red arrows represent the spin direction of Cr atoms. The crystal cells used in the calculation are also shown in Figure. }
\end{figure}

We now evaluate the magnetic ground state of  single-layer CrX$_3$.  Four possible different magnetic configurations as shown in Fig.~\ref{fig:magnetic} are considered\cite{xiao_di}. Our calculations indicate that the most stable magnetic state is the FM state. This  means that intra-ferromagnetism in the bulk form are retained upon exfoliation. The magnetic moment of a single unit cell is equal to 3 $\mu_B$ accurately and the local magnetic moments of Cr in  CrX$_3$ calculated using the PBE (HSE06) functional are 2.871 (2.900), 2.937 (3.021), 3.012 (3.127), and 3.103 (3.305) $\mu_B$ . This suggests that single layer CrX$_3$ are robust intrinsic ferromagnetic materials with large magnetic moments.

Magnetic anisotropy energy (MAE) is another technologically important parameter for  magneto-electronic applications. Ferromagnetic materials with small MAE will result in super-paramagnetic rather than ferromagnetic behavior. We have thus calculated the MAE of single-layer chromium trihalides using the PBE functional which includes the spin-orbit coupling (SOC). The results as shown in Table.~\ref{tab:monolayer} indicate that easy axis is along \emph{c}-direction for all CrX$_3$ compounds and isotropy holds in the basal plane. The corresponding MAE is  120, 32,186, and 686 $\mu$eV per Cr atom for CrF$_3$, CrCl$_3$, CrBr$_3$ and  CrI$_3$, respectively. Bulk CrI$_3$ was also reported to have the strongest magnetic anisotropy among chromium trihalides, \cite{crI3} which is also in accordance with our prediction.  It should be pointed out that the spins of the Cr atoms are found to lie in the basal plane in bulk CrCl$_3$, whereas bulk CrBr$_3$ and CrI$_3$ have an easy axis along the \emph{c}-direction. To confirm the easy axis of single-layer CrCl$_3$, we have also calculated magnetic anisotropy energy of single-layer CrCl$_3$ using HSE06 functional.  Our HSE06 results indicate that the easy axis of single-layer CrCl$_3$ is \emph{c}-axis with a magnetic anisotropy energy 13.5 $\mu$eV per Cr atom, which is in consistent with our PBE calculations.  The transition from an easy-plane to an easy-axis for  CrCl$_3$ upon exfoliation could be understood by interplay between crystal field splitting and spin-orbital coupling, which is beyond the scope of the present work. We also find that the MAE of  CrBr$_3$ and  CrI$_3$ are much larger than those of commonly used ferromagnetic materials such as bulk Fe (1.4 $\mu$eV per atom), and Ni (2.7 $\mu$eV per atom) \cite{mae_fe}, and are also comparable to those of Ca$^{2+}$ doped La(Mn,Zn)AsO  (509 $\mu$eV per atom) \cite{LaMnAsO}. The large MAE of single-layer CrBr$_3$ and  CrI$_3$ predicted here makes them promising candidates for low-dimensional magneto-electronic applications.

 In order to calculate the magnetic exchange interactions, we considered four different magnetic configurations as shown in Fig.~\ref{fig:magnetic}. Using  Eq.~(\ref{eq1}), the magnetic energy  of two CrX$_3$ formulas can be explicitly expressed as:
\begin{equation}
E_{FM/N\acute{e}el}=E_0-(\pm 3J_1+6J_2\pm 3J_3) |\vec{S}|^2
\end{equation}
\begin{equation}
E_{zigzag/stripy}=E_0-(\pm J_1-2J_2\mp3J_3) |\vec{S}|^2
\end{equation}
We then obtain exchange couplings by comparing with DFT results and also the Curie temperature using MC simulations. As listed in Table \ref{tab:mag_ml}, both the nearest neighbour ($NN$)  exchange interaction $J_1$ and the next nearest neighbour ($NNN$) interaction  $J_2$ are ferromagnetic, but the third nearest neighbour ($3NN$) interaction $J_3$ is anti-ferromagnetic.  Both the magnitudes of the magnetic interaction and the Curie temperature increase from F to I.   To evaluate the importance of $J_2$ and $J_3$, we also calculate  $J_{NN}$  only considering the $NN$ interactions.  Our calculations indicate that $J_{NN}$  is slightly smaller than $J_1$. However, it should be noted that the magnitude of $J_2$ is one order larger than  $J_3$  and the weight coefficient of $J_2$ in the Hamiltonian is twice that of $J_1$ in the Ferromagnetic state. In particular, for CrI$_3$, we find that the ratio of the contribution of $J_2$ and  $J_1$ in the Hamiltonian as evaluated by $\frac{2\times J_2}{J_1}$ is about 0.45, which implies the importance of  $J_2$  in the accurate determination of  the Curie temperature.  Using $J_{NN}$,T$_C$ of chromium trihalides are estimated to 36 ,36 ,51 and 56~K  respectively. The corresponding experimental Curie temperature of  bulk CrBr$_3$  \cite{jsps}and CrI$_3$ \cite{crI3} are  37 K and 61 K, respectively. When  the next  and third nearest neighbour exchange interactions are included, the T$_C$  increases to 41, 49, 73 and 95 K respectively. To evaluate the accuracy of the PBE calculations for magnetic exchange interactions, we also calculate the  exchange interaction only considering the $NN$ interaction using HSE06, which is also listed in the Table.~\ref{tab:mag_ml} for comparison. The results given by HSE06 turn out to be very close to the PBE except the CrF$_3$, which indicates the validity of the present PBE calculations.
\begin{table}
\centering
\caption{\label{tab:mag_ml}The exchange coupling and Curie temperature of single-layer chromium trihalides.}
\begin{ruledtabular}
\begin{tabular}{ccccccccccc}
	&$J_{NN}^{\rm HSE06}$	&$J_{NN}$	&	$J_1$	&$J_2$	&	$J_3$&	T$_C^{NN}$	&T$_C$	\\
	&meV&meV	&meV	&meV	&	meV&	K	&K	\\
\hline
CrF$_3$	&	0.8688	&	1.7904	&	1.8028	&	0.0672	&	-0.0128	&	36	&	41	\\
CrCl$_3$&	1.8728	&	1.7908	&	1.9232	&	0.2264	&	-0.1328	&	36	&	49	\\
CrBr$_3$&	2.4216	&	2.4524	&	2.5988	&	0.3800	&	-0.1464	&	51	&	73	\\
CrI$_3$	&	3.3216	&	2.7096	&	2.8624	&	0.6384	&	-0.1532	&	56	&	95	\\
\end{tabular}
\end{ruledtabular}
\end{table}

\begin{figure}
\centering
\includegraphics[width=0.45\textwidth]{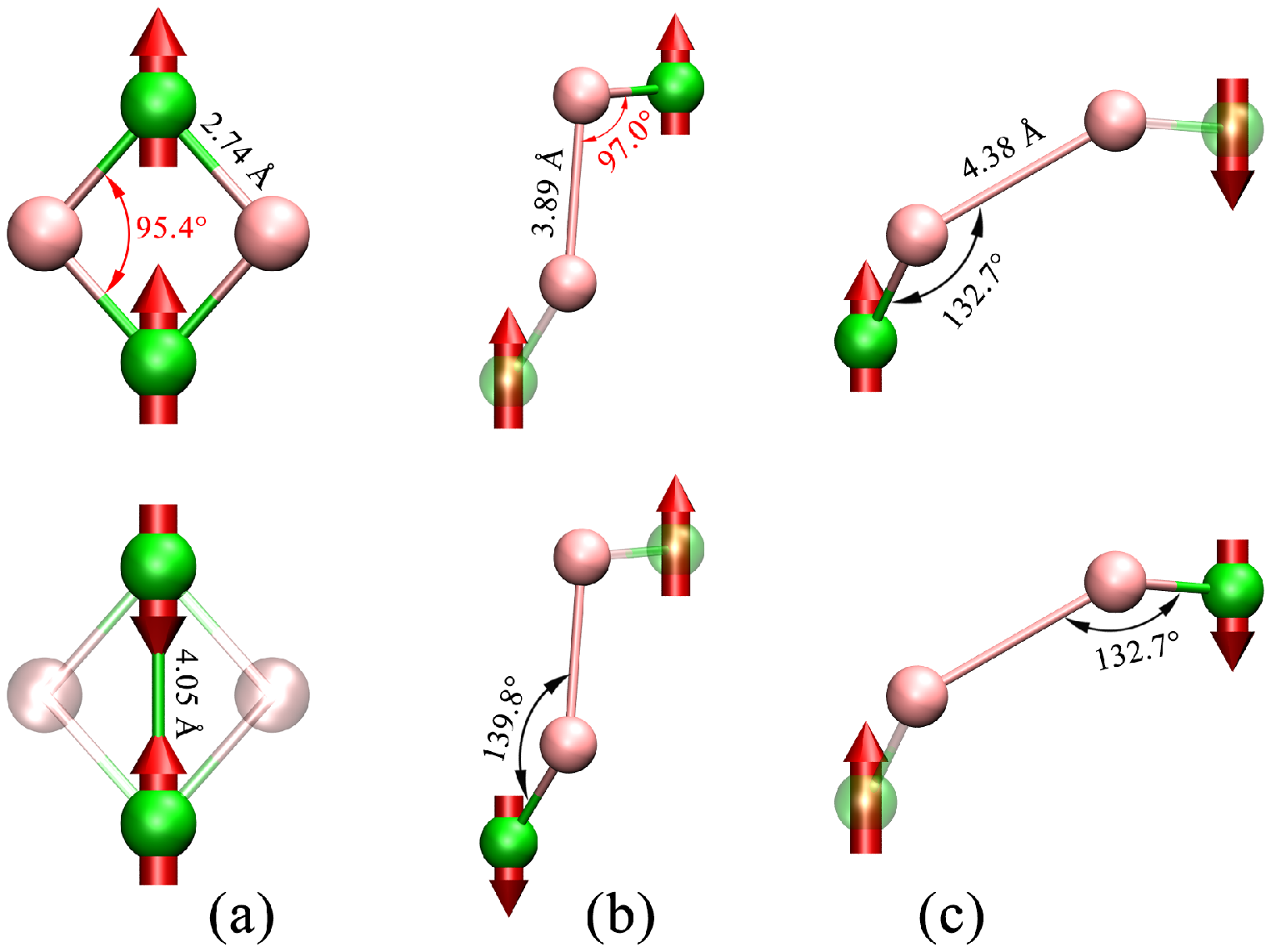}
\caption{\label{fig:path}(Color online) Possible paths for the mediation of NN (a), NNN (b) and 3NN (c) magnetic exchange interactions in ferromagnetic  single-layer CrI$_3$. (Anti-)Parallel arrangement of the Cr$^{3+}$ ions represent the  (anti-)ferromagnetic exchange interaction. }
\end{figure}
Now, we turn our attention to the microscopic origin of the exchange interactions in the single-layer chromium trihalides. The NN interaction  $J_1$  can be well understood from the competition between the direct exchange of the Cr-Cr sites and the superexchange mediated through the X ions. \cite{xiao_di,pla}. As shown above, the electronic configuration of Cr$^{3+}$ is $t_{2g}^3e_g^0$. The corresponding direct exchange will be weekly antiferromagnetic \cite{good}. Taking CrI$_3$ as an example, the Cr-X-Cr angle (95.4$^\circ$) is close to 90$^\circ$ in superexchange path shown in Fig.~\ref{fig:path}-(a), which is expected to be ferromagnetic. Due to the large distance (4.05 \AA) between the two Cr ions, which reduces direct antiferromagnetic exchange, the FM superexchange will dominate the total exchange interaction.   The underlying mechanism of $J_2$ and $J_3$ is more subtle. The cations are separated by two anions in the possible superexchange path, forming an extended cation-anion-anion-cation interaction path. Goodenough argued  \cite{good} that the rules for an extended  cation-anion-anion-cation path are the same as those for cation-anion-cation interactions.  The extended  cation-anion-anion-cation interaction can thus be seen as the sum of exchange interactions from two cation-anion-anion "superexchange path". In addition, according to the Goodenough-Kanamori-Anderson rule  \cite{good} which has been used widely to understand the signs and strengths of superexchange interactions, a 180$^\circ$ superexchange  is antiferromagnetic while a 90$^\circ$  superexchange interaction is ferromagnetic. For an intermediate angle between 90$^\circ$ and 180$^\circ$ ,  FM and AFM will compete with each other and a crossover angle from FM to AFM superexchange should exist. Subramanian \emph{et al.} \cite{prl_angle} have investigated  the FM-AFM transformation by  controlling the cation-anion-cation angle in cuprate perovskite and find a crossover angle about 127$\pm$0.6$^\circ$. Thus, we can reasonably assume that  a large cation-anion-anion angle ( $>$130$^\circ$ ) contributes a AFM interaction while an angle near 90$^\circ$ presents a FM contribution. As suggested by Sivadas \emph{et al.}\cite{xiao_di}, there are many possible exchange paths for $J_2$ and $J_3$ . Since the shortest distance between the X$^{-}$ pairs in different layers is larger than that between in-plane pairs, the most possible exchange path for  $J_2$  and $J_3$ should involve two X anions on the same plane. As shown in Fig.~\ref{fig:path}-(c), both cation-anion-anion angle for $J_3$ is 132.7$^\circ$ .  Thus, $J_3$  should be AFM due to both AFM contributions from the two cation-anion-anion superexchanges. However, for the NNN exchange interaction in Fig.~\ref{fig:path}-(b), the two  cation-anion-anion angles are  139.8$^\circ$  and  97.0$^\circ$, which contributes AFM and FM superexchange interactions, respectively. The sign of $J_2$  will depend on the competition of the FM and AFM superexchange interactions. In CrX$_3$, FM dominates over AFM and the resulting extended superexchange is FM. In contrast, for other systems such as MnPS$_3$ (Se$_3$ ) and  CrSi(Ge)Te$_3$, the total NNN exchange interaction with the extended superexchange path is AFM.\cite{xiao_di}
\begin{figure}
\centering
\includegraphics[width=0.45\textwidth]{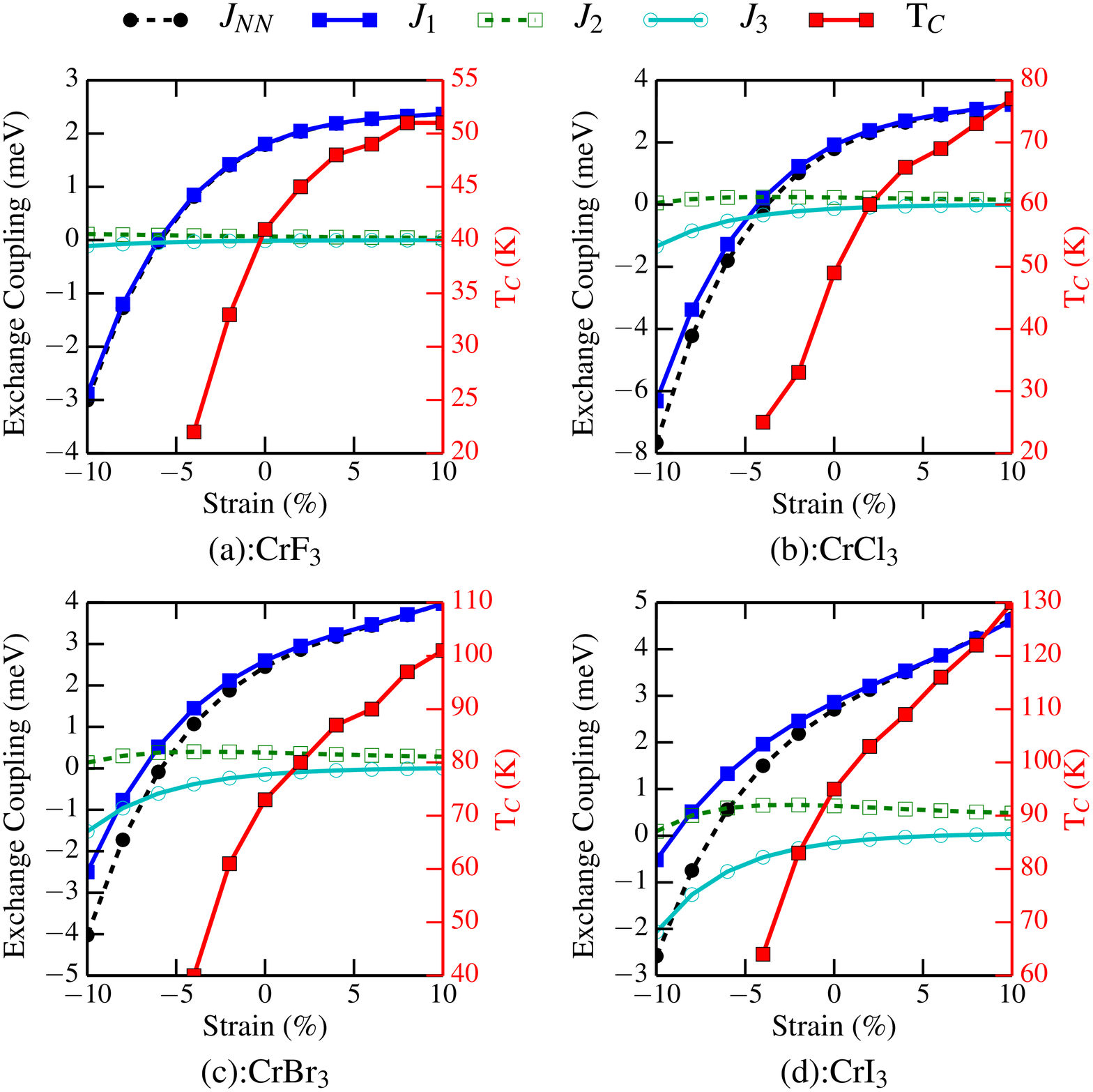}
\caption{\label{fig:strain}(Color online) The strain dependence of exchange interactions and Curie temperature of single-layer chromium trihalides.}
\end{figure}

Furthermore, we also investigate the magnetic interactions and Curie temperature  of CrX$_3$ under a biaxial strain. Results obtained based on interactions up to nearest-neighbours or third-nearest neighbours are  shown in Fig.~\ref{fig:strain}. We find that the NN and NNN exchange interactions  and the corresponding Curie temperature increase with strain, whereas the magnitude of 3NN interaction decrease. Yoshida \emph{et al.}\cite{Yoshida1997525} have investigated experimentally the pressure effect on the Curie temperature of CrBr$_3$ and found that T$_C$ decreases linearly with pressure, which agrees with our calculations.  The  Curie temperature can be tuned gradually with strain  and 130 K can be reached in the case of CrI$_3$ using a 10\% strain. Similar results are also reported in other systems such as CrSiTe$_3$.\cite{xiao_di,pla,jmc_li}  In addition, we also find that a compressive strain can also induce a FM to AFM transition. The  AF-N\'eel  phase becomes the most stable phase at a  strain of 6\%, 4\%, 6\%, 8\% for CrF$_3$, CrCl$_3$, CrBr$_3$ and CrI$_3$, respectively. Such a large compressive strain required to destroy the ferromagnetic state implies the robustness of  the intrinsic ferromagnetism in  single-layer CrX$_3$.
\section{\label{conclusion}Conclusions}
In summary, we have investigated the geometry, electronic, optical and magnetic properties of  chromium trihalides in single-layer and bulk forms. Our results indicate that the optB88-vdW functional can capture the interlayer vdW interaction of bulk CrX$_3$ quite well while HSE06 gives consistent electronic structures and optical properties with available experiment. Single-layer chromium trihalides are found to have very low cleavage energies and high in-plane stiffness, which suggests that free-standing single-layer chromium trihalides can be easily exfoliated from the bulk crystals.  The stability of single-layer chromium trihalides  is further confirmed using the phonon dispersions calculation and  \emph{ab initio} molecular dynamics simulations.  Furthermore,  our results also suggest that these 2D crystals exhibit robust long-range ferromagnetic order  with magnetic moments of 3$\mu_B$ per formula unit  and large magnetic anisotropy energies. The ferromagnetic nearest-neighbour exchange interactions can be attributed to the competition between direct antiferromagnetic exchange interactions and near 90$^\circ$ ferromagnetic superexchange interactions. The  next-nearest-neighbour FM exchange interactions can be understood by the angle-dependent extended superexchange Cr-X-X-Cr interaction. Monte Carlo simulations based on the classical Heisenberg model predicts Curie temperatures of chromium trihalides up to 90 K, and can be further increased by applying biaxial tensile strains. Electronic structure calculations indicate that both bulk and 2D chromium trihalides are half semiconductors with indirect band gaps and the absorption edges of CrBr$_3$ and CrI$_3$ lie in the visible range, which suggests that they are promising candidates for future low-dimensional semiconductor spintronic, magneto-electronic and magneto-optic applications.

\begin{acknowledgments}
The financial supports from the National Natural Science Foundation of China (Grant No. 11004018) and HK PolyU (Grant No. 1-ZE14) are appreciated gratefully. This work was also supported by the construct program of the key discipline in Hunan province and aid program for Science and Technology Innovative Research Team in Higher Educational Institutions of Hunan Province. Wei-Bing Zhang thanks Mr. Xingxing Li  in University of Science and Technology of China for useful discussions about magnetic exchange interaction.
\end{acknowledgments}

\end{document}